# The effect of the substrate on the Raman and photoluminescence emission of atomically thin MoS$_2$

*Michele Buscema[1], Gary .A. Steele[1], Herre S.J. van der Zant[1] and Andres Castellanos-Gomez[1]*

[1]Kavli Institute of Nanoscience, Delft University of Technology, Lorentzweg 1, 2628 CJ Delft, The Netherlands

**ABSTRACT**

We quantitatively study the Raman and photoluminescence (PL) emission from single layer molybdenum disulfide (MoS$_2$) on dielectric (SiO$_2$, hexagonal boron nitride, mica and the polymeric dielectric Gel-Film®) and conducting substrates (Au and few-layer graphene). We find that the substrate can affect the Raman and PL emission in a twofold manner. First, the absorption and emission intensities are strongly modulated by the constructive/destructive interference within the different substrates. Second, the position of the A$_{1g}$ Raman mode peak and the spectral weight between neutral and charged excitons in the PL spectra are modified by the substrate. We attribute this effect to substrate-induced changes in the doping level and in the decay rates of the excitonic transitions. Our results provide a method to quantitatively study the Raman and PL emission from MoS$_2$-based vertical heterostructures and represent the first step in *ad-hoc* tuning the PL emission of 1L MoS$_2$ by selecting the proper substrate.

## 1   Introduction

Layered transition metal dichalcogenides (TMDCs) are increasingly attracting interest due to their extraordinary properties in single-layer form. Within this class of materials, molybdenum disulfide (MoS$_2$) has a fairly high in-plane mobility [1, 2], a large Seebeck coefficient [3], remarkable mechanical properties [4-6] and a large and direct bandgap (single layer: 1.88 eV) [7, 8]. Thus, for optoelectronic applications, MoS$_2$ represents an interesting complement to graphene especially as photo-active material for the visible part of the spectrum [9, 10].

Owing to the large bandgap and exciton binding energy, single-layer $MoS_2$ shows strong photoluminescence (PL) emission [7, 8, 11]. The study of the PL properties of single layer $MoS_2$ has led to the discovery of interesting phenomena such as the control of charged excitons via electrostatic doping and optical control of valley population [11-15]. Samples of 1L $MoS_2$ on hexagonal boron nitride (h-BN) flakes were also studied because of their stronger PL emission as compared with 1L $MoS_2$ on regular SiO2/Si substrates. This suggest an important effect of the substrate on the luminescence properties of $MoS_2$, similarly to the effect of solvent environment reported by Mao *et al* [16].

In this work, we systematically study the effect of the substrate on the PL and Raman signal from single and few-layer $MoS_2$. We find a strong influence of the substrate on both the PL peak wavelength and intensity for the single layer $MoS_2$. On all studied substrates, single layer $MoS_2$ shows a factor ~4 enhancement of PL efficiency, relative to the commonly used $SiO_2$ substrate. This can be explained by a combined effect of the substrate on the doping in the $MoS_2$ and on the radiative decay rates of neutral and charged excitons.

## 2 Experimental

### 2.1 Sample fabrication and optical characterization setup

We prepare the studied $MoS_2$ flakes on the $SiO_2$ (285nm)/Si and Gel-Film® substrates by micromechanical exfoliation of natural $MoS_2$ (*SPI Supplies, 429ML-AB*) with blue Nitto tape (*Nitto Denko Co., SPV 224P*). The heterostructures of $MoS_2$ on few layer graphene (FLG), h-BN, mica and gold (Au) are prepared following the method developed in Refs [17, 18]. Briefly, we prepare the substrate of interest via mechanical exfoliation (FLG, h-BN, mica) or metal deposition (Au) on the same $SiO_2$/Si wafers. Then we exfoliate $MoS_2$ flakes on a flexible, transparent stamp. The stamp is rigidly connected to a glass slide, inverted and mounted into a modified micromanipulator (*Süss microtech*). Both the substrate and the stamp are then placed under an optical microscope with a long working-distance lens. This enables one to locate the region of interest on the sample (e.g. a FLG flake) and align the stamp carrying the selected $MoS_2$ flake over it. By carefully bringing the $MoS_2$-stamp in contact with the surface, it is possible to deterministically transfer the selected $MoS_2$ flake on the substrate.

The quantitative optical characterization of the $MoS_2$ flakes is carried out with an *Olympus BX 51* microscope equipped with a *Canon EOS 600D* digital camera. The number of layers is determined by Raman and Atomic Force Microscopy (AFM) imaging. The AFM (*Digital Instruments D3100* with standard cantilevers with spring constant of ~40 N m$^{-1}$ and tip curvature <20 nm) is used in amplitude modulation mode.

**2.2 Raman and photoluminescence spectroscopy setup**

Raman and photoluminescence (PL) spectra are recorded simultaneously in a micro-Raman spectrometer (*Renishaw in via*) in backscattering configuration excited with an Ar laser ($\lambda$ = 514 nm) as in Refs. [3, 19]. To reject the Rayleigh scattering, we employ a 50/50 beamsplitter and two notch filter centered at the laser line. The system is equipped with a single-pass spectrometer with a grating of 1800 grooves/mm and a Peltier-cooled CCD array. The slits are set to an aperture of ~20 μm providing a resolution of about 0.5 cm$^{-1}$. Typical integration times are in the order of 10 s and power in the order of 250 μW to avoid heating effects [8].

# 3   Results and discussion

We have studied the effect of both conducting and insulating substrates on the PL response of single layer $MoS_2$. As insulating substrates, we selected silicon oxide ($SiO_2$), Gel-Film®, hexagonal boron nitride (h-BN) flakes and muscovite mica flakes. As conducting substrates, we employed gold (Au) and few-layer graphene (FLG) flakes. Silicon oxide and gold have been selected because they are widely used in the literature and serve as reference insulating and conducting materials. Gel-Film® is a commercially available poly(dimethylsiloxane) derivative, appealing for its possible technological relevance as flexible transparent substrate for optoelectronic applications [20].

By employing h-BN, mica and FLG flakes as substrates one can study the effect of highly crystalline insulating and conducting substrates on the PL emission of $MoS_2$ and it constitutes an essential step towards the characterization of novel heterostructures based on h-BN, $MoS_2$ and graphene which are recently attracting increasing attention [11, 21-23]. On $SiO_2$, Gel-Film® and Au, atomically thin $MoS_2$ was deposited by micromechanical exfoliation from bulk natural molybdenum disulfide [24]. To deposit $MoS_2$ over h-BN, mica or FLG flakes, the h-BN, mica or FLG flakes are firstly transferred onto a $SiO_2$/Si wafer by mechanical exfoliation and then a deterministic all-dry stamping method was used to transfer the $MoS_2$ onto the selected h-BN or FLG flake [12, 13].

### 3.1 Characterization of the heterostructures

To reliably assign the number of $MoS_2$ layers deposited on each substrate, we used a combination of complementary microscopy techniques consisting of: optical microscopy (in reflection and transmission modes, if possible), atomic force microscopy and Raman and PL microscopy. The characterization of a $MoS_2$/FLG heterostructure is shown in Figure 1 as a representative example. For the characterization of $MoS_2$ on $SiO_2$, Gel-Film®, h-BN, Au and mica we refer the reader to the Electronic Supplementary Material (ESM). Figure 1a shows an optical micrograph of a few-layer MoS$_2$/FLG (on a SiO$_2$/Si substrate) heterostructure. For clarity, the contours of the FLG and MoS$_2$ flakes are outlined with white and orange dashed lines respectively. The optical contrast allows one to distinguish the MoS$_2$ flake from the FLG. Moreover, different number of MoS$_2$ layers on the FLG flake have different optical contrast, easing their optical identification as described in Refs [25-27].

Figure 1b shows a sketch of the experimental setup employed to record the emission spectra of the MoS$_2$ samples in broad range of wavelngths (for details, see ESM). We can acquire both the Raman and PL part of the emission spectrum at the same time. This allows one to accurately correlate the PL emission and Raman properties. Characteristic combined Raman and PL spectra for 1, 2, 3 and 4 layers of MoS$_2$ on FLG are plotted in Figure 1c. All the spectra in Figure 1c show Raman features from MoS$_2$ close to 520 nm (labeled with an asterisk), the silicon peak at ~528 nm, then the G and 2D Raman active modes of FLG and PL signatures of the MoS$_2$ flake at ~ 630 nm and ~670 nm associated with emission from the B and A excitonic species, respectively [7, 8, 11, 28].

We now focus on the Raman part of the spectrum, plotted in Figure 1d. The measured points (open circles) are well fitted by Lorentzian functions (solid lines). The in-plane (E$^1_{2g}$) and out-of-plane (A$_{1g}$) Raman modes are clearly visible and their frequency changes with the number of layers. The difference between the E$^1_{2g}$ and A$_{1g}$ modes ($\Delta f$) is known to steadily increase with the number of layers and, therefore, it is a reliable quantity to count the number of layers of MoS$_2$ on SiO$_2$ [5, 7, 8, 25, 29]. Figure 1e plots the measured $\Delta f$ for MoS$_2$ on FLG as a function of the layer number and shows its continuous increase. We find a similar relation between the $\Delta f$ and the number of MoS$_2$ layers for the other studied substrates (see ESM). This confirms that $\Delta f$ is an appropriate quantity to assign the number of MoS$_2$ layers on a variety of substrates.

Turning our attention to the PL part of the spectra, we see that both the wavelength and intensity of the maximum PL emission are dependent on the number of MoS$_2$ layers. To clarify this, we plot the peak intensity (Figure 1f) and peak wavelength (Figure 1g) of the PL emission as a function of the number of MoS$_2$ layers on the FLG substrate. The intensity drops and the peak wavelength is strongly red-shifted when the number of layers increases. This is associated with the known direct-to-indirect bandgap transition as the number of layer changes from 1 to 2 or more [7, 8].

### 3.2 Photoluminescence spatial characterization

Raman and PL mapping is a commonly used tool to characterize TMDCs and graphene [8, 19, 30, 31]. To spatially resolve the Raman and PL emission, spectra are collected while rastering the sample under the microscope objective with a step size of ~ 300 nm. Figure 2 shows the combined Raman and PL maps obtained on the MoS$_2$/FLG heterostrucure presented in Figure 1 (similar measurements for other substrates are presented in the ESM). Figure 2a shows a zoom-in of the MoS$_2$/FLG heterostructure of Figure 1a with the previously determined number of layers. The scanned region is delimited by the black box drawn on the MoS$_2$ flake in Figure 2a. The edge of the FLG flake is represented by the dashed line and is determined by mapping the intensity of the graphene G-peak, simultaneously acquired with the MoS$_2$ Raman and PL spectra (see ESM).

Figure 2b shows a colormap representing the position of the A$_{1g}$ Raman active mode. Darker regions (low A$_{1g}$ frequency) in the MoS$_2$ flake correspond to 1L MoS$_2$ layers, while lighter regions (high A$_{1g}$ frequency) are the thicker part of the flake (2L and 3L), in agreement with previous works [29]. Figure 2c and Figure 2d show a color map of the integrated PL intensity and peak wavelength, respectively. The single layer region shows the highest emission intensity and shortest emission wavlength due to its direct bandgap [7, 8]. This is consistent with the known change of the MoS$_2$ band structure as the number of layers decreases [7, 32]. We systematically performed the measurements shown in Figure 2 for the MoS$_2$ flakes deposited on different substrates obtaining qualitatively similar results (see ESM). Both Raman and PL features are homogeneous over the surface of the flakes. This indicates that the all-dry transfer process does not introduce Raman or PL active defects in the MoS$_2$ flakes.

### 3.3 Modelling the substrate-dependent interference effects.

When deposited onto a substrate, the MoS$_2$ flakes and the substrate can be visualized as a vertically stacked medium with several internal interfaces (Figure 3a). The differences in the optical constants and thickness of each component of the medium give rise to optical interference in both the incoming and the emitted light. This substrate-induced interference will have an effect on the absorption of the excitation light as well as the Raman and PL intensity. For the Raman emission of both MoS$_2$ [33] and graphene [30-32] on a standard SiO$_2$/Si substrate, a model to treat optical interferences has already been developed.

In this section, we extend the model used in Refs [33], [30-32] to also include the effect of optical iterference at the typical PL emission wavelengths of MoS$_2$ laying on a multi-layered substrate. The main aim of the model is to calculate the total absorption and emission intensity by taking into account multiple internal reflection at every interface between the media composing the vertical heterostructure (Figure 3a). Note that this approximation is well proven to capture the main experimental features [33, 34]. We employed both the effective medium approach and the transfer matrix formalism to model the interference effects. Both approaches delivered the same result, supporting our methodology. The full derivations for the model and expression for the electric field amplitudes can be found in the ESM.

We calculate the total emission intensity for all substrates geometries and for a freely suspended 1L MoS$_2$. We then define a substrate-dependent enhancement factor of the following form $\Gamma^{-1} = \frac{I_{MoS_2}^{\text{freestanding}}}{I_{MoS_2}^{\text{on substrate}}}$ where $I_{MoS_2}^{\text{freestanding}}$ is the emission intensity from a freestanding 1L MoS$_2$ and $I_{MoS_2}^{\text{on substrate}}$ is the total emission intensity from 1L MoS$_2$ on a substrate. The results of the calculation of $\Gamma^{-1}$ for 15 nm FLG and 285 nm SiO$_2$ are plotted in Figure 3b. For the other substrates, we refer the reader to the ESM. The enhancement factor can reach values larger than 10, it is nonlinear with the wavelength and highly dependent on the substrate.

We normalize the measured data to the freestanding condition by multiplying by $\Gamma^{-1}$. Through this normalization procedure, it is possible to attribute the differences among the spectra to intrinsic differences in the 1L MoS$_2$ induced by the different substrates. Figure 3c plots the spectra for 1L MoS$_2$ over FLG before (black solid line) and after correction (light blue solid line) and Figure 3d shows the zoom in the Raman part of the spectra (the dots are measured data and the solid lines are Lorentzian fits). After normalization, there is a factor ~8 increase in the PL emission and a factor ~6 in the Raman intensity. Note that, despite a strong increase in the intensity, there is no change in the peak position, neither for the Raman nor for the PL.

Figures 3e(f) shows the full spectra (Raman part) for 1L MoS$_2$ on 285nm SiO$_2$ before and after correction. In this case, the effect of the normalization is less than in the case of FLG, in agreement with a much lower $\Gamma^{-1}$ value. This difference stems from the difference in height and optical constants between the two substrates. Again, the position of the peaks is unaffected by the interference effect of the substrate. This holds for all studied substrates (see ESM). We can therefore conclude that the substrate-dependent interferometric situation does not perturb the peak emission wavelength while strongly influences the intensity of the emission.

### 3.4 Effect of the substrate on the Raman

Raman spectroscopy has proven to be an effective tool to determine, not only the number of layers of MoS$_2$, but also the built-in strain [35] in the layers as well as their doping level [16, 36]. Therefore, analyzing the Raman part of the spectrum allows one to further characterize the fabricated MoS$_2$ structures. Figure 4a shows a comparison of the Raman spectra measured for MoS$_2$ single layers deposited onto the different substrates. The intensity $E^1_{2g}$ and $A_{1g}$ modes is clearly modulated by the substrate. We then look at the effect of the substrate on the frequency at which these modes occur.

Figure 4b plots the measured frequency of the $E^1_{2g}$ mode on every substrate. The position of the $E^1_{2g}$ peak seems rather insensitive to the substrate material since it shows less than ~0.4 cm$^{-1}$ variation across the different substrates. The $E^1_{2g}$ mode is known to be sensitive to the strain in the material [35]. Rice *et al.* [35] measured a shift in the $E^1_{2g}$ mode of 2.1 cm$^{-1}$ per % of uniaxial strain and Hui *et al.* [37] measured a shift of 4.7 cm$^{-1}$ per % of biaxial strain. In the present case, we can then estimate a maximum strain level of ~0.2% in case of uniaxial strain and of ~0.09% in case of biaxial strain. Furthermore, the Raman response is homogeneous on the surface of the flakes (Figure 2 and Figures S1-S5). Thus, we can conclude that strain does not play a major role in our measurements.

While the $E^1_{2g}$ mode is barely affected by the substrate, the $A_{1g}$ mode shows a sizeable stiffening up to ~2 cm$^{-1}$ (Figure 4c). We note that for the measurements on mica, a spectral overlap between the $MoS_2$ $A_{1g}$ mode and the $B_g$ mode of bulk mica could arise [38]. We observe an increase in the full width at half maximum of the $MoS_2$ $A_{1g}$ mode on mica that could be related to this spectral overlap. We measure the lowest frequency for 1L $MoS_2$ on $SiO_2$ and the highest for h-BN.

A stiffening of the $A_{1g}$ mode can be associated with reduced electron density in 1L $MoS_2$ [16, 36]. For dielectric layers, doping can come from charged impurities at the substrate/$MoS_2$ interface [39]. While the $SiO_2$ substrate is known to have a high degree of charge impurities that result in high doping level, h-BN flakes and polymeric dielectrics have much lower density of charge impurities and, therefore, could induce a much lower doping level, in agreement with recent studies [2, 39, 40]. For Au and FLG substrates, the main doping mechanism could be direct charge transfer. Physisorbed molecules should not play a major role in the doping, since all the samples were fabricated in and exposed to the same environmental conditions and they were not annealed [40].

Another possible explanation for the stiffening of the $A_{1g}$ mode is a change in the strength of the dipolar interaction between the $MoS_2$ layer and the fixed charges in the different substrates. Since the $A_{1g}$ mode is the out-of-plane motion of the negatively charged S atoms with respect to the Mo atoms, a different electrostatic environment will provide a change in the potential landscape where the motion takes place. This will not induce intrinsic doping in the 1L $MoS_2$ but will affect the $A_{1g}$ frequency. By considering only the Raman part of our measurements, it is not possible to distinguish between the two effects. More insight on this, however, can be obtained by also considering the PL spectrum on different substrates.

### 3.5 Effect of the substrate on the PL emission.

Photoluminescence from monolayers TMDCs has been extensively studied and it proved to be a valuable tool to gain insight in intrinsic material properties [7, 8, 11, 13, 41]. In this section, we study the effect of the substrate on the PL emission of 1L $MoS_2$. Figure 5a plots the PL spectra for 1L $MoS_2$ on the different substrates. The spectra in Figure 5a show the A and B excitonic peaks at ~ 655nm and ~ 630 nm respectively. Moreover, another common feature at ~670 nm appears. This feature can be associated with the emission from charged excitons (trions) of the A excitonic transition (A⁻) [11]. The other peaks are substrate-dependent features identified according to Refs [28, 42]. The dashed black line connects the positions of the peak of the PL intensity across all spectra. Both the wavelength and intensity of the maximum PL emission are dependent on the substrate.

By fitting the data to Lorentzian functions (see ESM), we extract quantitative information about the PL spectra. First we discuss the PL peak intensity. Figure 5b plots the maximum PL intensity measured for all substrates. 1L MoS$_2$ on SiO$_2$ shows the lowest PL intensity while all the other studied substrates provide roughly the same enhancement of the PL emission. It is noteworthy that a flexible polymeric substrate results in a similar PL enhancement to that of samples on h-BN with a much easier fabrication route. From studies on carbon nanotubes, it is known that SiO$_2$ can reduce PL emission by scattering with surface optical phonons [43-45]. Metallic substrates (Au and FLG) can also affect the PL intensity via additional non-radiative paths for exciton recombination (such as charge transfer processes and dipole-dipole interaction) [46-50].

Figure 5c plots the PL emission peak wavelength for all substrates. There is a large ($\geq 10$ nm) blue-shift in the emission peak for Au, Gel-Film ®, FLG and h-BN as compared to SiO$_2$ and mica. This blue-shift can be explained by a relative increase in the luminescence emission from neutral excitons. Due to the large exciton binding energy (~ 30 meV [11]), neutral excitons emit at a significantly higher energy than charged excitons. Therefore, an increase in the population of neutral excitons will result in a blue-shifted PL spectra [11, 40]. Thus, the measured blue-shift in the PL emission peak is an indication that the substrate affects the relative population of neutral to charged excitons.

The relative concentration of charged to neutral excitons can be related to their weight in the PL spectra, defined as:

$$\gamma = \frac{I_{A^-}}{I_A + I_{A^-}} = \frac{N_{A^-} \cdot \tau_{A^-}}{N_A \cdot \tau_A + N_{A^-} \cdot \tau_{A^-}}$$

where $I_i$ is the emission intensity, $N_i$ is the concentration and $\tau_i$ is the radiative decay rate of either the A or the A⁻ excitonic species. We can obtain $\gamma$ from the measured PL spectra by determining the relative PL emission intensity of the A and A⁻ transition (see ESM).

It is well known that in 2D electron systems the relative population of charged and neutral excitonic species is related to the doping level through a chemical equilibrium of the form [11, 51, 52]:

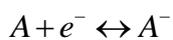

$$A + e^- \leftrightarrow A^-$$

Charged excitons (A⁻) are produced by the binding of a free electron (e⁻) to a neutral exciton (A). Thus the doping level in $MoS_2$ can affect the relative population of charged to neutral excitons. Doping is also known to affect the $A_{1g}$ Raman active mode [36]. It is therefore possible to correlate a change in the trion spectral weight with the frequency of the $A_{1g}$ mode.

In Figure 5c, we plot the trion spectral weight against the frequency of the $A_{1g}$ mode for 1L $MoS_2$ on the studied substrates. The gray dashed line corresponds to the calculated $\gamma$ including the change in doping as a function of the frequency of the $A_{1g}$ mode [36] and decay rates from Ref [11] (see ESM for further details). The data collected on $SiO_2$, Gel-Film ®, Au and FLG are in good agreement with the model. This indicates that for these substrates we can explain a change in $\gamma$ with a substrate-induced reduction in the doping level in the 1L $MoS_2$. The data points on h-BN and mica are not in good agreement with the model. For these points, also a change in the decay rates of neutral and charged excitons should be included: for 1L $MoS_2$ on h-BN it appears that the decay rate of neutral excitons is much larger than for 1L on $SiO_2$ and while the opposite seems to be the case for 1L $MoS_2$ on mica. This indicates that the substrate can affect both the background doping and the radiative decay rates of excitonic transition.

## 4    Conclusions

In summary, we have systematically studied the Raman and PL properties of 1L $MoS_2$ transferred on several substrates. While the substrate has little to no influence on the $E^1_{2g}$ Raman mode, it largely effects the $A_{1g}$ mode. This suggests that the substrate do not cause a noticeable amount of strain in the 1L $MoS_2$ flakes and indicates a possible effect on the doping level. The study of the photoluminescence reveals that, compared to $SiO_2$, all the measured substrates provide an enhanced emission. Interestingly, flexible polymeric substrates show larger enhancements than $MoS_2$/h-BN heterostructures with a much simpler fabrication route and the possibility of applications in flexible transparent electronics. Strickingly, the substrate induces sizable changes in the peak emission wavelength and intensity. These changes can be related to substrate-induced variations in the spectral weight and radiative decay rate of charged and neutral excitons.

**Acknowledgements**

The authors thank Lihao Han and Arno Smets for fruitful discussion about the photoluminescence setup. This research was supported by the Dutch organization for Fundamental Research on Matter (FOM) and a Marie-Curie Fellowship.

**Electronic Supplementary Material**: Supplementary material include experimental methods and details. Optical, AFM and Raman characterization of the measured devices, Explanation of the observed shifts of the Raman peaks, Dataset of Raman spectra of 1, 2, 3 and 4 L $MoS_2$ on dielectric ($SiO_2$, Gel-Film®, h-BN) and metallic (Au, FLG) substrates, Dataset of PL spectra of 1, 2, 3 and 4 L $MoS_2$ on dielectric ($SiO_2$, Gel-Film®, h-BN) and metallic (Au, FLG) substrates, Spatially resolved PL maps, Spatially resolved FLG G-peak.

**FIGURES.**

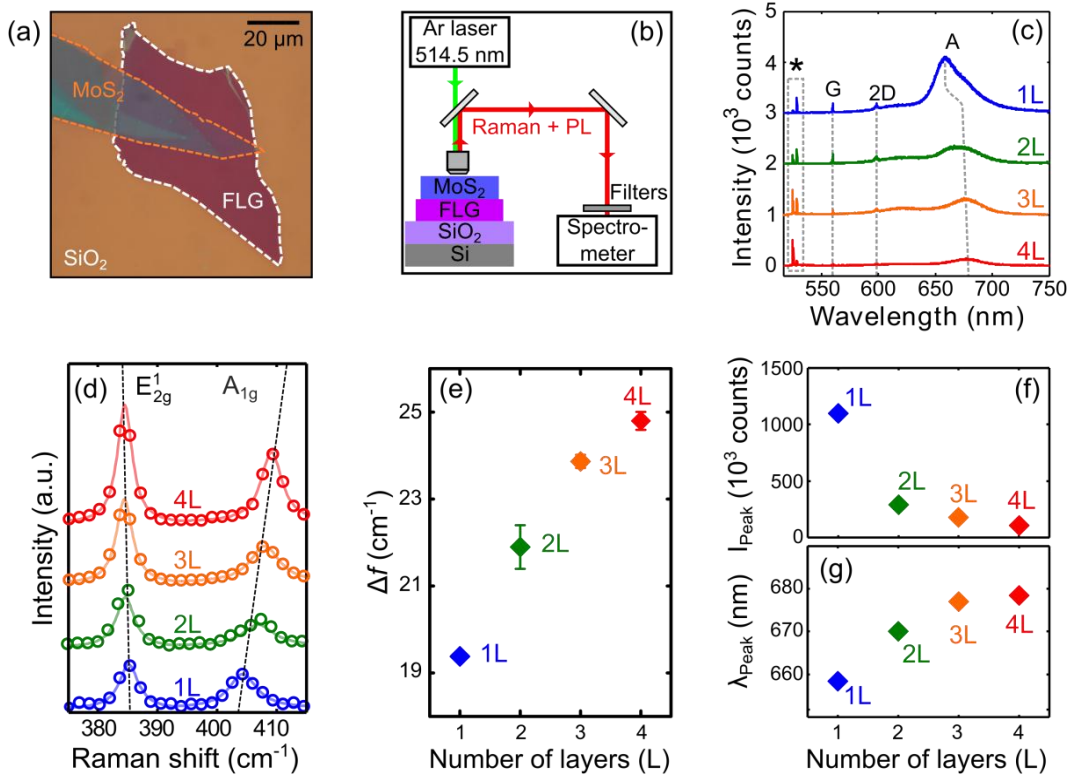

**Figure 1.** Optical and Raman characterization of the fabricated heterostructures (a) Optical micrograph of one of the studied heterostructures: the MoS$_2$ flake is on top of a FLG flake which is supported by a SiO$_2$ (285nm)/Si substrate. The contours of the FLG and MoS$_2$ flakes are outlined for clarity. (b) Schematics of the experimental setup. The heterostrucure, as in panel (a) – out of scale for clarity), is illuminated with 514.5 nm light through a 100X (NA = 0.95) objective. The Raman and photoluminescence signal is collected by the same objective and sent to a spectrometer after filtering. (c) Raman and PL spectra of 1, 2, 3 and 4 layer MoS$_2$ on top of FLG showing the Raman-active modes of MoS$_2$ (gray box labelled with an asterisk), the Raman active G and D modes of FLG and the B and A excitonic species of MoS$_2$. The spectra are shifted vertically for clarity. The dashed gray lines are guides to the eyes. (d) Raman spectra of 1, 2, 3 and 4 layer MoS$_2$ on top of FLG showing the $E^1_{2g}$ and the $A_{1g}$ MoS$_2$ Raman-active modes. The spectra are shifted vertically for clarity. The open circles are the experimental data and the shaded solid lines are Lorentzian fit to the data. The dashed black lines are guides to the eye. (e) Frequency difference between the $E^1_{2g}$ and the $A_{1g}$ modes as a function of the number of MoS$_2$ layers for MoS$_2$/FLG heterostructures. (f) PL peak intensity as a function of the MoS$_2$ number of layers on FLG (g) PL peak wavelength as a function of the MoS$_2$ number of layers on FLG.

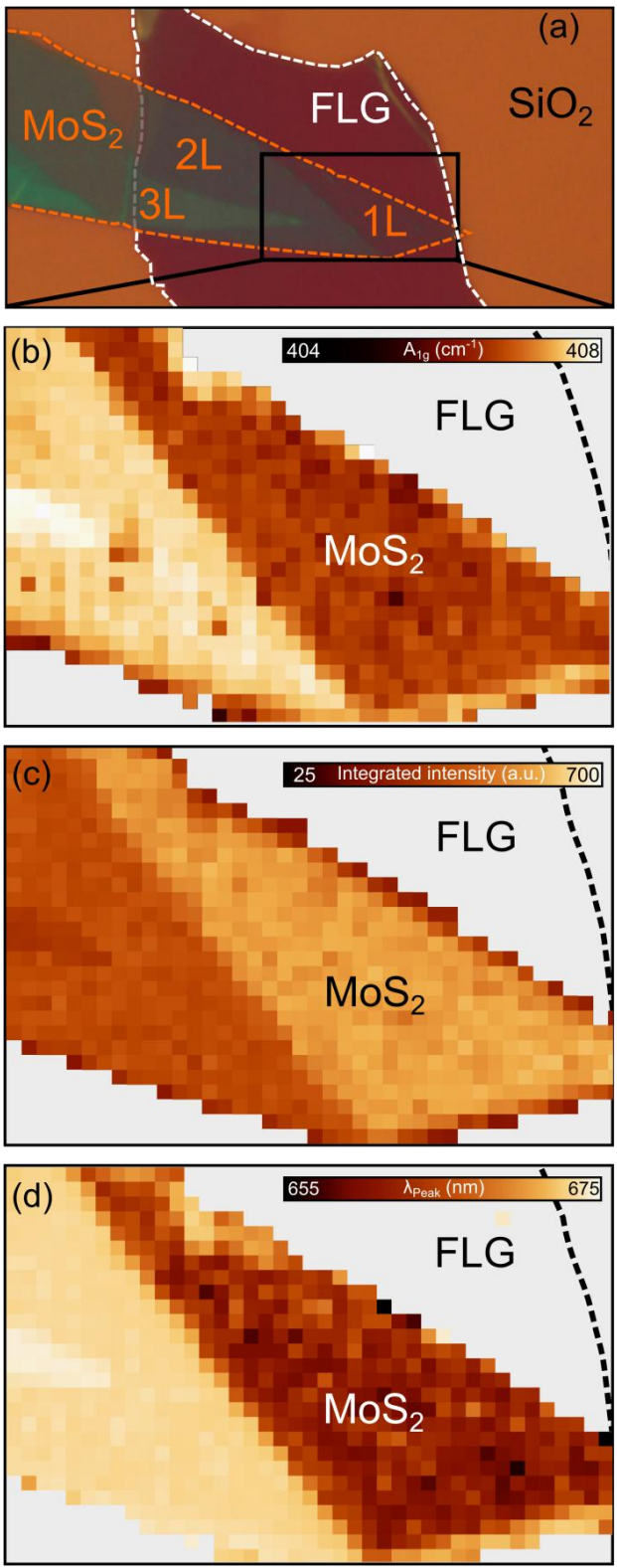

**Figure 2**. Spatial maps of the substrate dependent Raman and PL emission of single- and few-layer MoS$_2$. (a) Zoom of the optical micrograph in Figure 1a. The dark rectangle is the area where the Raman and PL spectra were recorded. (b) Color map of the frequency position of the A$_{1g}$ mode. (c) Color map of the integrated PL intensity. (d) Color map of the PL wavelength peak. Note that in panels b, c and d the gray background represents area were no peak was found. In panels b, c and d the dashed lines identify the edge of the FLG flake. This edge is determined by superposition of a color map of the intensity of the G-peak taken at the same time in the same region (see ESM).

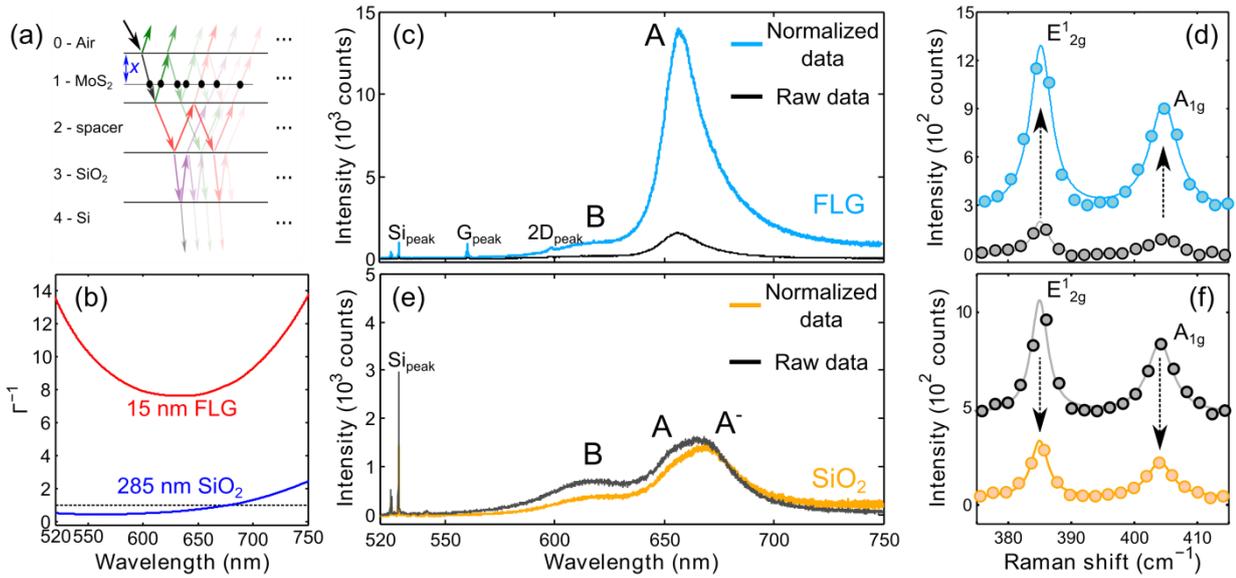

**Figure 3**. Effect of substrate-induced optical interference on combined Raman and photoluminescence spectra of 1L MoS$_2$ on FLG and SiO$_2$. (a) Schematic geometry and ray paths used to model the interference effects. The intensity of the arrow represents, schematically, the absorption due to the various substrates. The black dots inside the MoS$_2$ layer represent the point of absorption/emission. The different layers are numbered for convenience. Spacer stands for either FLG, h-BN, Au and Mica. (b) Enhancement factor ($\Gamma^{-1}$) *vs* wavelength for 15nm FLG and 285nm SiO$_2$. The dashed black line indicates $\Gamma^{-1} = 1$. For $\Gamma^{-1} < 1$ the intensity of the signal is suppressed, for $\Gamma^{-1} > 1$ the signal is enhanced (c) Combined Raman and PL spectra of 1L MoS$_2$ on FLG before (black solid line) and after normalization (light blue solid line) for the interference effects. Peaks are identified in the plot. (d) Raman part of the spectra shown in panel c: the black dots represent the measured data before normalization and the gray solid line is a double Lorentzian fit to the data before normalization; the light blue dots and solid line represent measured data and Lorentzian fit after normalization for interference effects. The arrows schematically visualize the effect of the normalization. (e) Combined Raman and PL spectra of 1L MoS$_2$ on SiO$_2$ before (black solid line) and after normalization (orange solid line) for the interference effects. Peaks are identified in the plot. (f) Raman part of the spectra shown in panel a: the black dots represent the measured data before normalization and the gray solid line is a double Lorentzian fit to the data before normalization; the orange dots and solid line represent measured data and Lorentzian fit after normalization for interference effects. The arrows schematically represent the effect of the normalization.

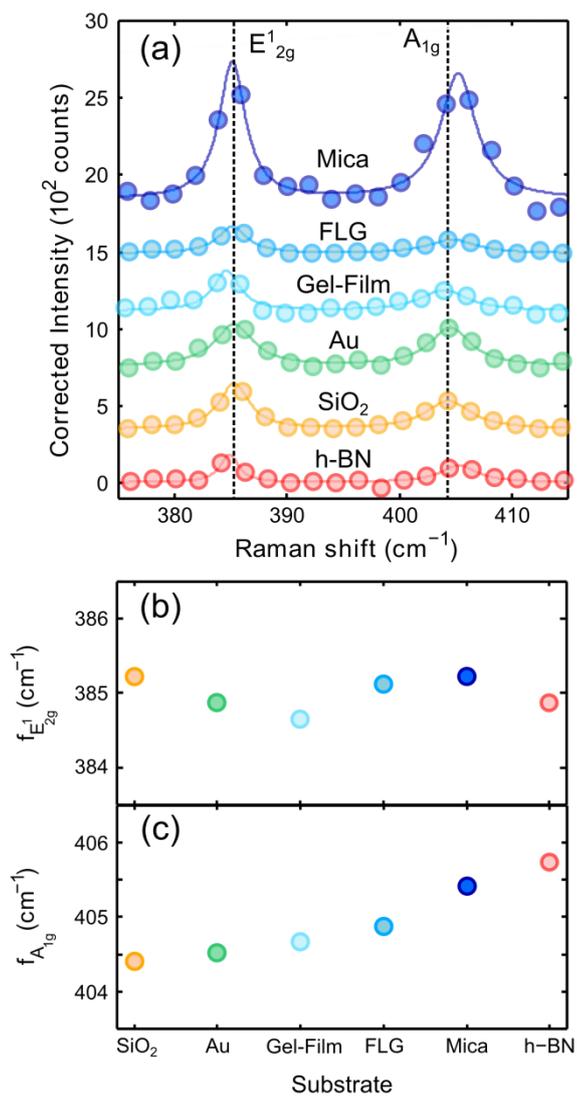

**Figure 4**. Effect of the substrate on the Raman modes of 1L $MoS_2$ (a) Normalized Raman spectra for 1L $MoS_2$ on Mica, FLG, Gel-Film®, Au, $SiO_2$ and h-BN (shifted vertically for clarity). The dots are the experimental points; the solid lines are Lorentzian fits. The dashed solid lines corresponds to the $E^1_{2g}$ and $A_{1g}$ frequency on $SiO_2$. (b) Frequency of the $E^1_{2g}$ (upper panel) and the $A_{1g}$ (lower panel) Raman-active modes as a function of substrate material.

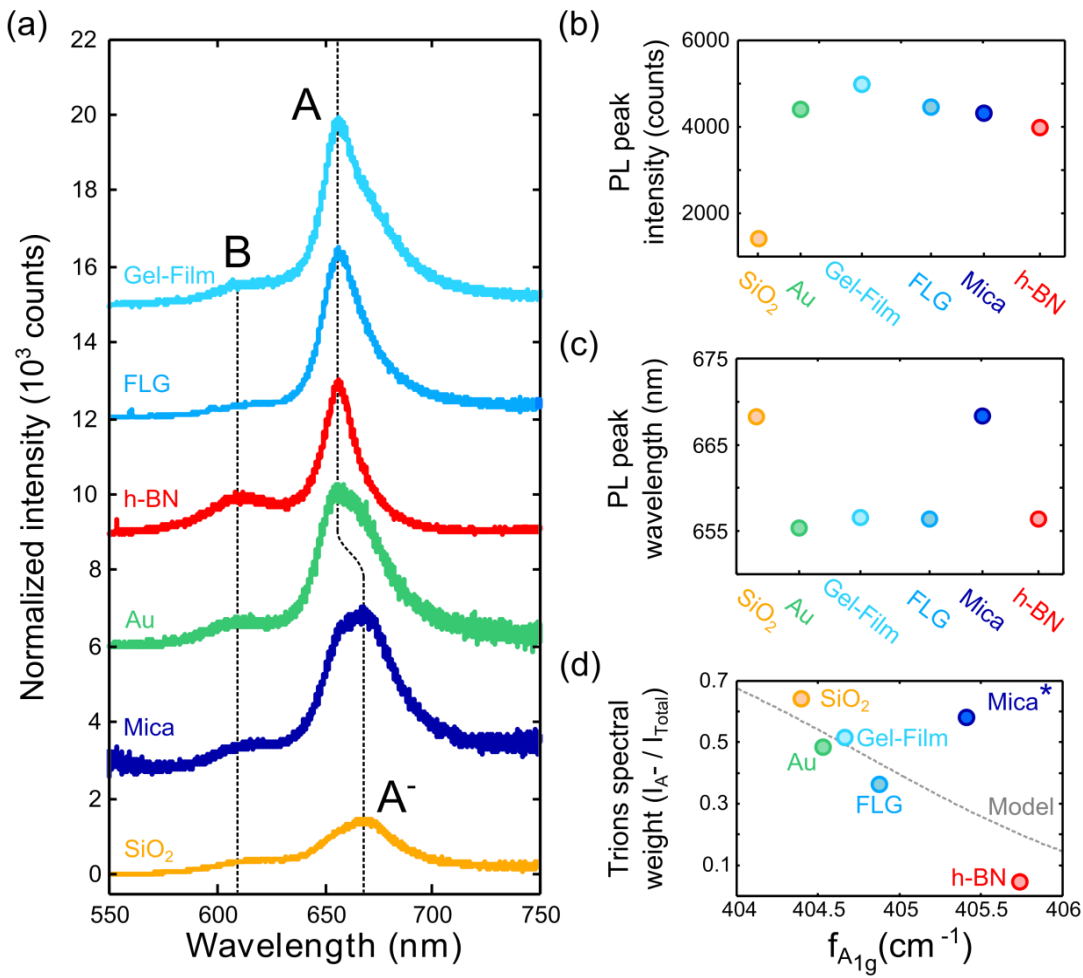

**Figure 5.** Effect of the substrate on the photoluminescence of 1L MoS$_2$. (a) Photoluminescence spectra of 1L MoS$_2$ on Gel-Film® (light blue solid line), FLG (blue solid line), h-BN (red solid line), Au (green solid line), Mica (dark blue solid line), SiO$_2$ (orange solid line). Spectra are corrected for interference effects and are shifted vertically for clarity. The dashed black lines indicate the position of peak emission, shifting from the A (neutral) to the A$^-$ (charged) excitonic transition and the position of the B excitonic transition. (b) Peak PL emission wavelength as a function of the frequency of the A$_{1g}$ mode for different substrates. (c) Maximum PL intensity as a function of the frequency of the A$_{1g}$ mode for different substrates. (d) Trions spectral weight (defined as ratio of the A- integrated intensities *vs* the sum of the A and A$^-$ integrated intensities) as a function of the frequency of the A$_{1g}$ MoS$_2$ mode for different substrates. The gray dashed line is the result of the mass action model for trions. The star on the measurement on mica in panel d indicates that the accuracy of the determination of the A$_{1g}$ mode frequency for 1L MoS$_2$ might be affected by the presence of a mica Raman-active mode at ~405 cm$^{-1}$.

# Electronic Supplementary Material

# The effect of the substrate on the Raman and photoluminescence emission of atomically thin MoS$_2$

*Michele Buscema[1], Gary .A. Steele[1], Herre S.J. van der Zant[1] and Andres Castellanos-Gomez[1]*

[1]Kavli Institute of Nanoscience, Delft University of Technology, Lorentzweg 1, 2628 CJ Delft, The Netherlands

**List of supplemental materials**

1) Experimental methods and fabrication details.

2) Optical, AFM and Raman characterization of the measured devices.

3) Dataset for Raman on 1,2,3 and 4 MoS$_2$ layers on the various substrates.

4) Derivation of 5 media interference model.

5) Effect of spacer height on emission intensity.

6) Fit for 1L MoS2 measurements

7) Mass action model for trion in PL emission.

8) Model calculations.

9) Spatially resolved PL maps.

10) Determination of the FLG edge.

**Experimental methods and details**

*Preparation and optical characterization setup*

We prepare the studied structures on the SiO$_2$ (285nm)/Si and Gel-Film® substrates by micromechanical exfoliation of natural MoS$_2$ (*SPI Supplies, 429ML-AB*) with blue Nitto tape (*Nitto Denko Co., SPV 224P*). The heterostructures are prepared following the method developed in Ref [1]. Briefly, we prepare the substrate of interest via mechanical exfoliation (Few layer graphene, hexagonal boron nitride and mica) on the same SiO$_2$/Si wafers. Then we exfoliate MoS$_2$ flakes on a flexible, transparent stamp. The stamp is rigidly connected to a glass slide, inverted and mounted into a modified micromanipulator (Süss microtech). Both the substrate and the stamp are then placed under an optical microscope with a long working-distance lens. This enables us to locate the region of interest on the sample (e.g. a few layer graphene – FLG – flake) and, at the same time, align the stamp carrying the selected MoS$_2$ flake. By carefully bringing the MoS$_2$-stamp in contact with the surface, it is possible to deterministically transfer the selected MoS$_2$ flake on the substrate.

The quantitative optical characterization of the MoS$_2$ flakes is carried out with an *Olympus BX 51* microscope equipped with a *Canon EOS 600D* digital camera (see Figure 1a in the main text and Figures S1-S6). The number of layers is determined by Raman and Atomic Force Microscopy (AFM) imaging (see figures S1-S5). The AFM (*Digital Instruments D3100* with standard cantilevers with spring constant of ~40 N m$^{-1}$ and tip curvature <20 nm) is used in amplitude modulation mode to measure the topography and to determine the number of MoS$_2$ layers flakes.

*Raman and photoluminescence spectroscopy setup*

Raman and photoluminescence (PL) spectra are recorder simultaneously in a micro-Raman spectrometer (*Renishaw in via*) in backscattering configuration excited with an Ar laser (λ = 514.5 nm) as in Refs. [2, 3]. To reject the Rayleigh scattering, we employ a 50/50 beamsplitter and two notch filter centered at 514 nm. Typical integration times are in the order of 10 s and power in the order of ~250 µW to avoid heating effects [4]. The system is equipped with a single-pass spectrometer with a grating of 1800 grooves/mm (in first-order angular position) and a Peltier-cooled CCD array. The slits are set to an aperture of ~20 µm. The system resolution is about ~ 0.5 cm$^{-1}$.

**Optical, AFM and Raman characterization of the measured samples**

In Figure S1-S6, we show the identification and layer counting procedure for a representative selection of the studied MoS$_2$-based structures. In Figures S1-S5, we present optical micrographs of the studied structures and the insets show the optical contrast calculated according to ref [5]. We also present amplitude modulation AFM micrographs of the studied devices and the linecuts show the height values along the indicated lines. The height data are in fair agreement with literature values [6] except for the case of Gel-Film® which might be affected by the difference between the MoS$_2$-tip and substrate-tip interaction. presence of absorbates on the surface and/or by the different amplitude setpoint used during AFM measurements [7]. Lastly, we present the result of spatially resolved Raman measurements: we show the difference ( $\Delta f$ ) between the frequency of the two most prominent Raman peaks (E$^1_{2g}$ and A$_{1g}$). This procedure allows us to reliably assign the number of MoS$_2$ layers present in the studied area.

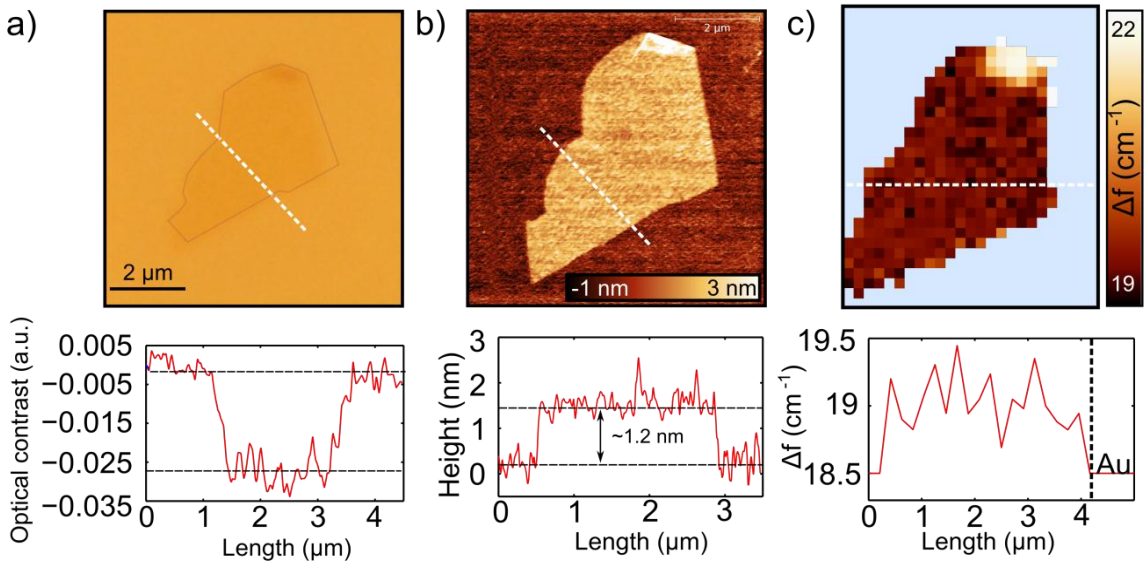

**Figure S1.** (a) Optical micrograph of 1L MoS$_2$ over a gold surface. Inset shows the calculated optical contrast. (b) AFM micrograph of the area in panel (a). The inset shows the height values along the white dashed line. (c) Color map of $\Delta f$ of the studied device. The inset shows the values $\Delta f$ along the white dashed line.

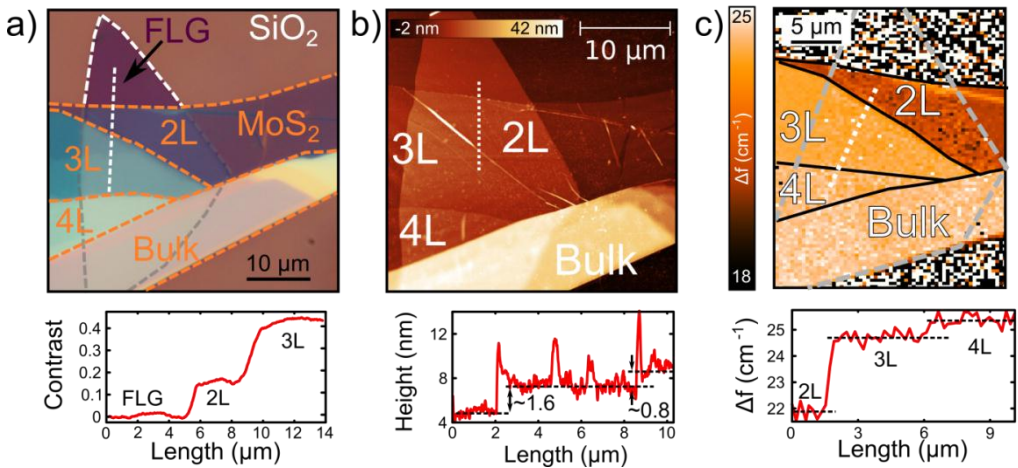

**Figure S2.** (a) Optical micrograph of a few-layer MoS$_2$ flake over a few-layer graphene (FLG) flake supported on a SiO$_2$/Si substrate. The contours of the MoS$_2$ (FLG) have been highlighted with orange (white/gray) dashed lines to facilitate identification. Inset shows the calculated optical contrast. (b) AFM micrograph of the area in panel (a). The inset shows the height values along the white dashed line. (c) Color map of $\Delta f$ of the studied device. The edges of the FLG flakes are highlighted with gray dashed lines and have been identified by superposition of the intensity of the G peak. The inset shows the values of $\Delta f$ along the white dashed line.

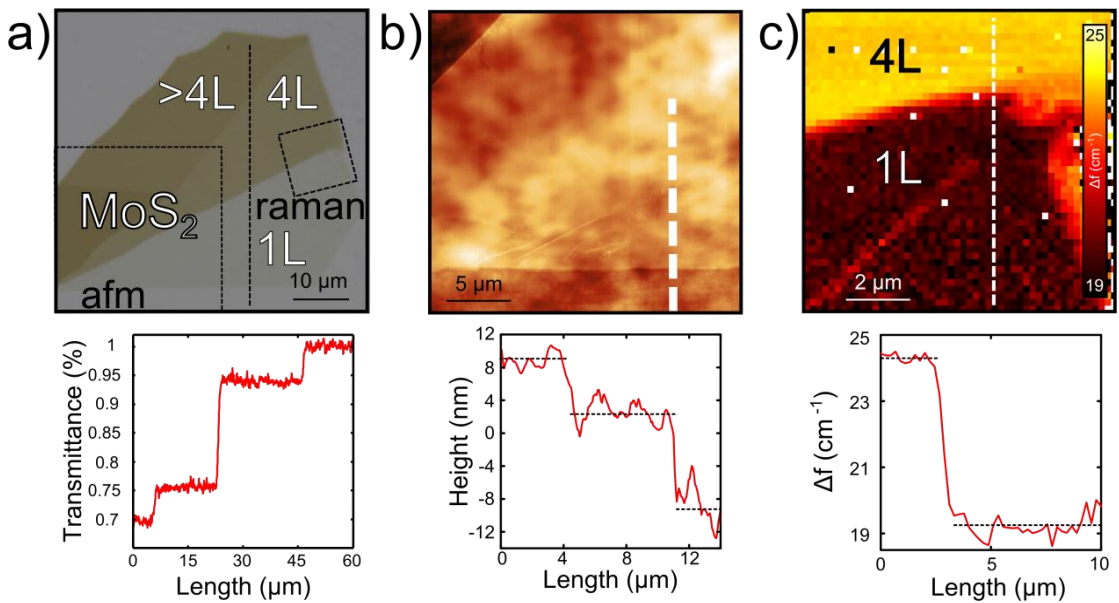

**Figure S3.** (a) Optical micrograph of a few-layer MoS$_2$ flake over Gel-Film® surface. Inset shows the calculated transmittance along the solid black line. (b) AFM micrograph of the area indicated in panel (a). The inset shows the height values along the white dashed line. The height values are not in agreement with literature due difficulties in acquiring the AFM topography. These difficulties stem from the elastomeric nature of the substrate and are unavoidable. (c) Color map of Δ$f$ of the studied device measured in the area indicated in panel (a). The inset shows the values of Δ$f$ along the white dashed line.

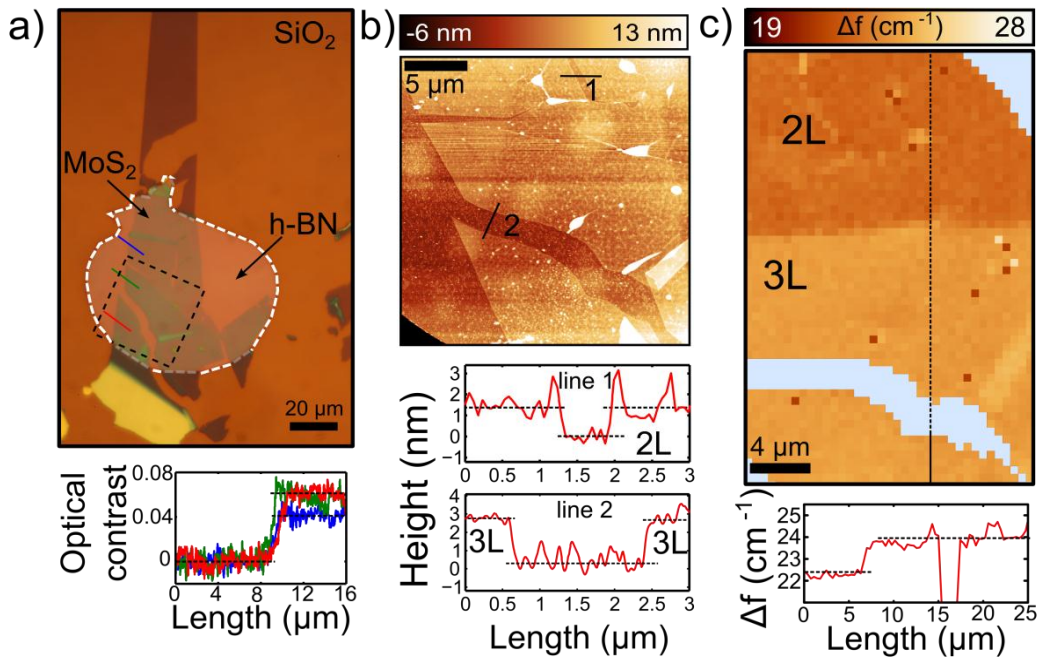

**Figure S4.** (a) Optical micrograph of a few-layer MoS$_2$ flake over h-BN flake supported on a SiO$_2$/Si substrate. The contour of the h-BN flake is highlighted by dashed lines to facilitate its identification. Inset shows the calculated optical contrast along the solid red, green and blue lines. (b) AFM micrograph of the area indicated

in panel (a). The inset shows the height values along the black solid dashed lines. (c) Color map of $\Delta f$ of the studied device measured in the area indicated in panel (a). The inset shows the values of $\Delta f$ along the black solid line.

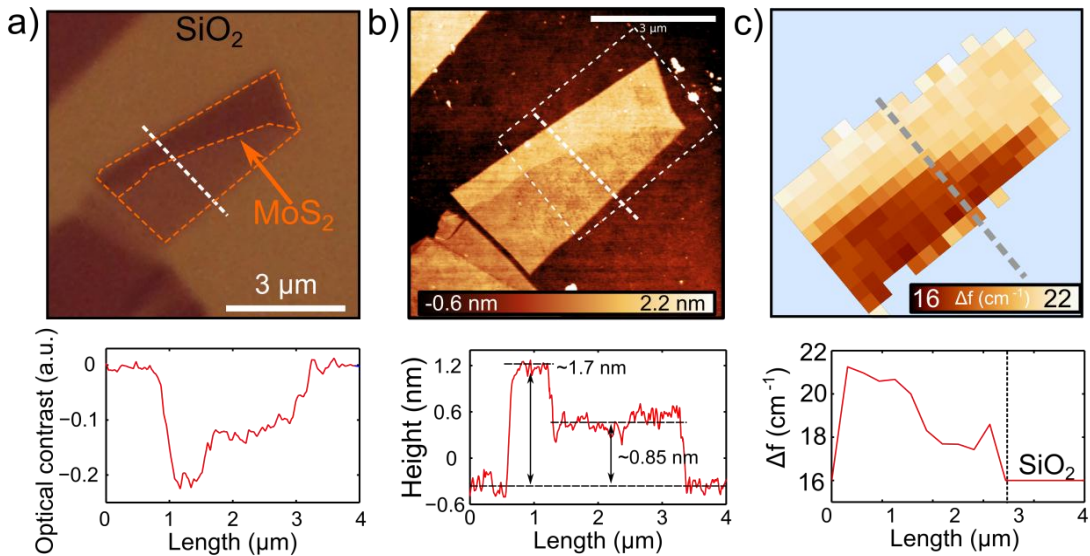

**Figure S5.** (a) Optical micrograph of a few-layer MoS$_2$ flake over a SiO$_2$/Si substrate. The contour of the MoS$_2$ flake is highlighted by dashed lines to facilitate its identification. Inset shows the calculated optical contrast along the dashed white line. (b) AFM micrograph of the area indicated in panel (a). The inset shows the height values along the white dashed lines. (c) Color map of $\Delta f$ of the studied device measured in the area indicated in panel (b). The inset shows the values of $\Delta f$ along the gray dashed line.

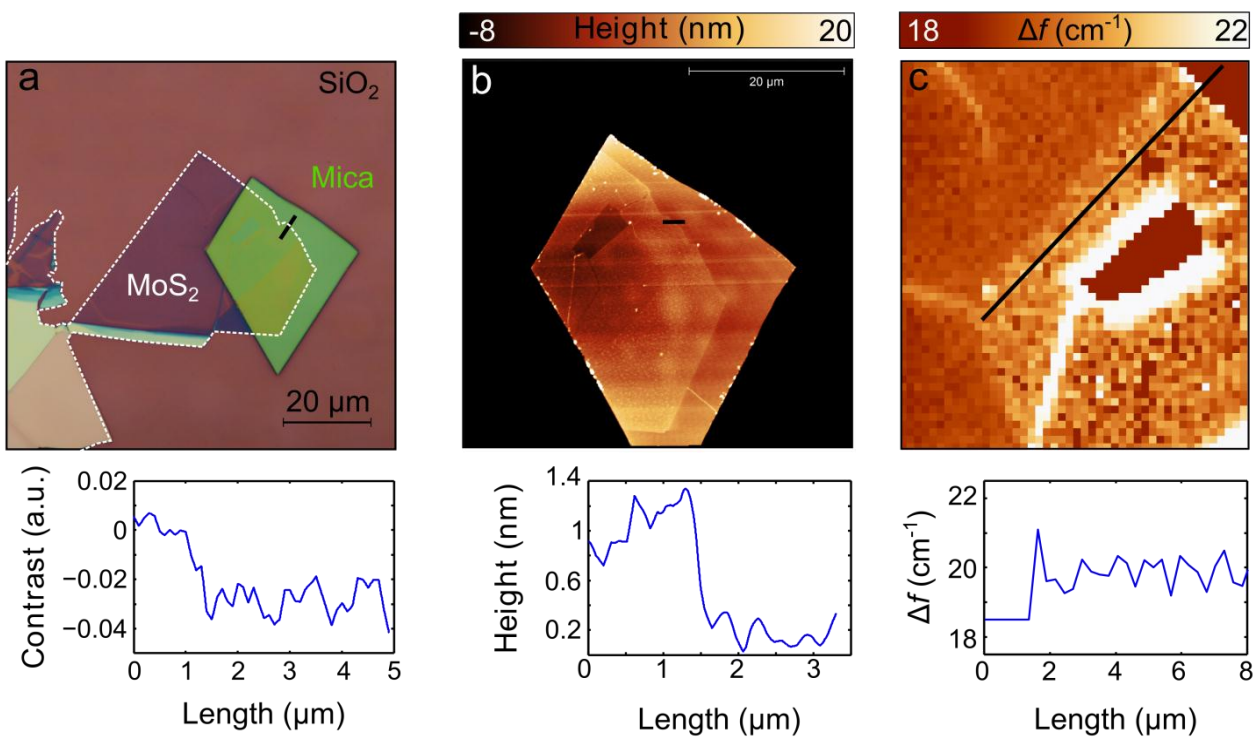

**Figure S6.** (a) Optical micrograph of a 1L MoS$_2$ flake over a mica flake. The contour of the MoS$_2$ flake is highlighted by dashed lines to facilitate its identification. Inset shows the calculated optical contrast along the black solid line. (b) AFM micrograph of panel (a). The inset shows the height values along the black solid lines. (c) Color map of $\Delta f$ of the studied device (b). The inset shows the values of $\Delta f$ along the solid black line.

**Dataset of Raman spectra of 1, 2, 3 and 4 L MoS$_2$ on dielectric (SiO$_2$, Gel-Film®, h-BN) and metallic (Au, FLG) substrates.**

In this section, we show the complete dataset for the Raman measurement of single and few-layer MoS$_2$ over dielectric and metallic substrates.

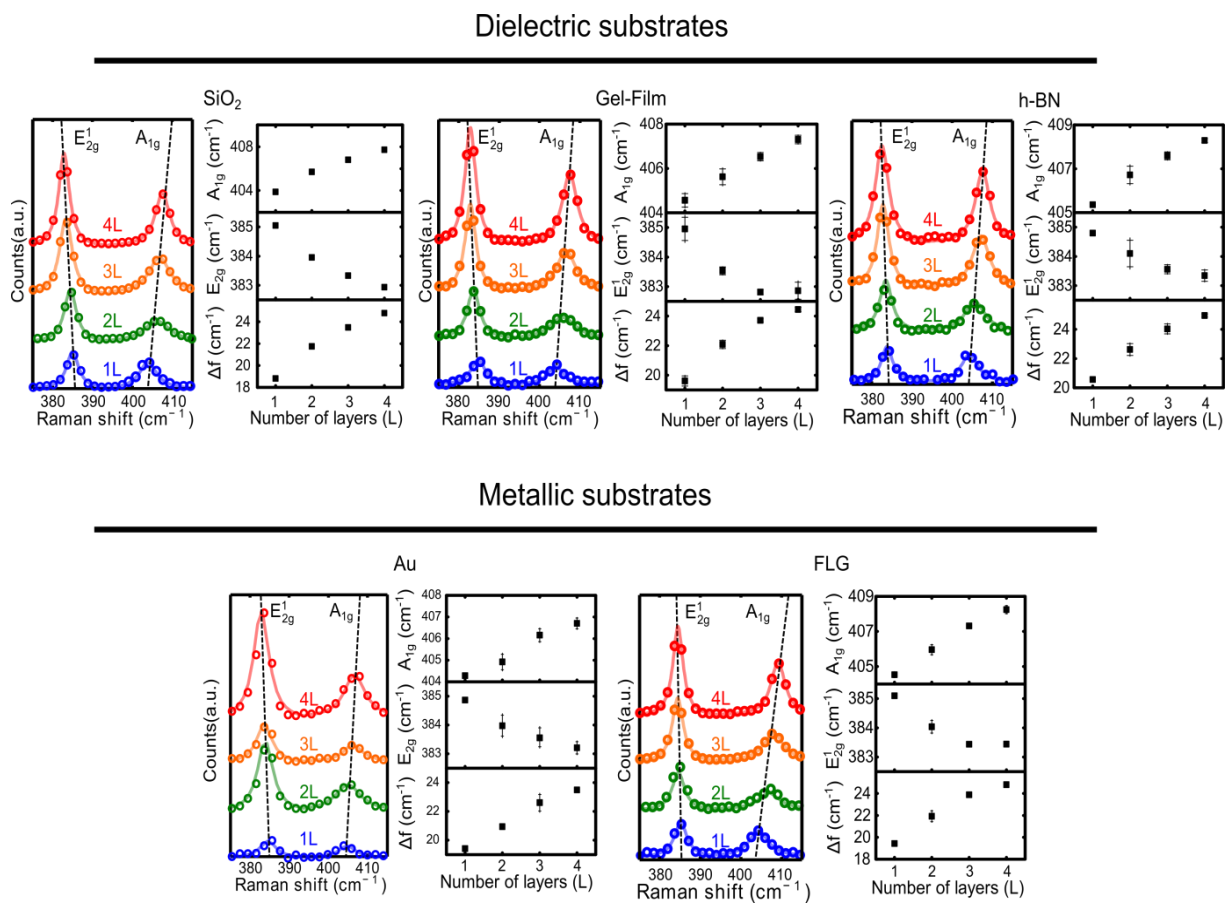

**Figure S7.** Raman spectra, frequency of the A$_{1g}$ and E$^1_{2g}$ modes and difference between the frequency of the aforementioned modes ($\Delta f$) for dielectric (SiO$_2$, Gel-Film® and h-BN) and metallic (Au, FLG) substrates

To easily compare the values on different substrate, we summarize the dataset in Figure S8. Figure S8a shows the frequency of the A$_{1g}$ and the E$^1_{2g}$ modes as a function of the number of MoS$_2$ layers and the substrate materials. Figure S8b plots the difference between the A$_{1g}$ and E$^1_{2g}$ frequency ($\Delta f$) as a function of the number of layers and for different substrate materials. The effect of the substrate is rather limited for the $E^1_{2g}$ mode, while is significant on the $A_{1g}$ mode also for 2, 3 and 4 layer MoS$_2$. The effect of the substrate on $\Delta f$ is less and less important for thicker MoS$_2$ flakes. This could be due to the screening of the substrate-generated electric field by the MoS$_2$ layers. Nonetheless, in the case of Au, the effect of the substrate is still pronounced which might be due to a stronger charge transfer.

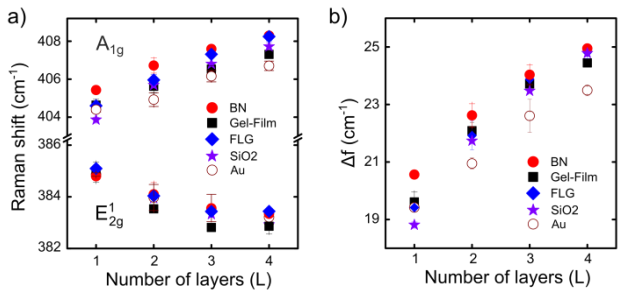

**Figure S8.** (a) Frequency of the $A_{1g}$ and $E^1_{2g}$ modes as a function of the number of MoS$_2$ layers for the different studied substrate. (b) Frequency difference between ($\Delta f$) between the $A_{1g}$ and $E^1_{2g}$ modes as a function of the number of MoS$_2$ layers for the studied substrates.

**Dataset of PL spectra of 1, 2, 3 and 4 L MoS$_2$ on dielectric (SiO$_2$, Gel-Film®, h-BN) and metallic (Au) substrates.**

In this section, we show the complete dataset for the PL measurement of single and few-layer MoS$_2$ over dielectric and metallic substrates. Note that this is raw data.

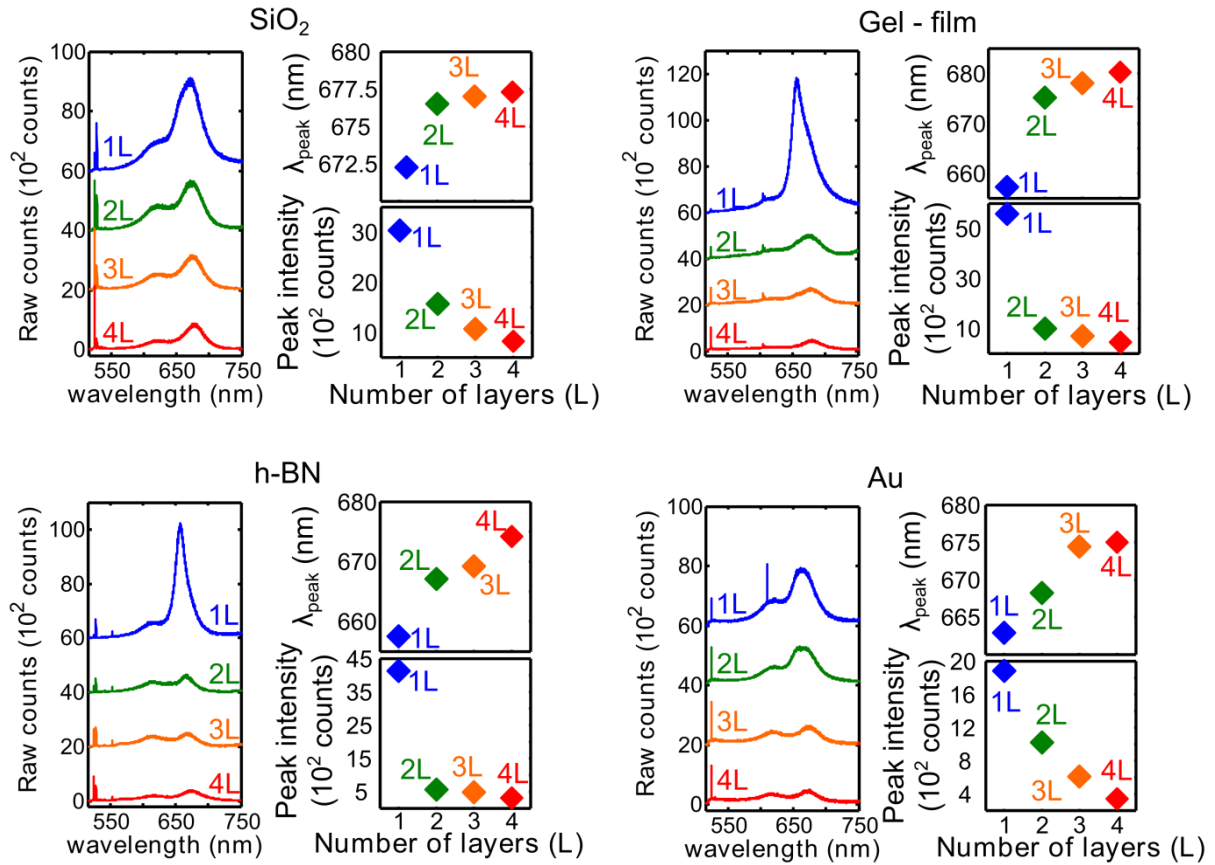

**Figure S8.** PL spectra, peak wavelength and peak emission intensity for dielectric (SiO$_2$, Gel-Film® and h-BN) and metallic (Au, FLG) substrates

**Derivation of 5 media interference model.**

In this section we present the derivation of the optical interference model for 5 media separated by 4 interfaces.

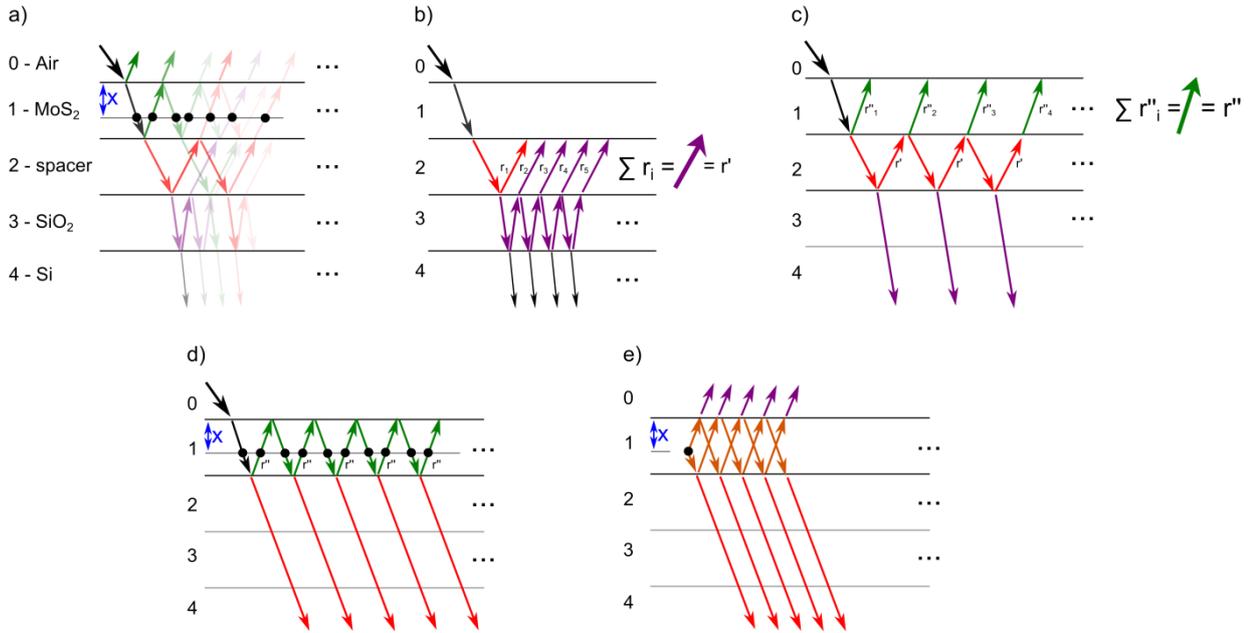

**Figure S9.** (a) Schematic ray diagram of the 5 media system, including multiple reflections at every interface and absorption and emission at a depth $x$ in the MoS$_2$ layer (dots). Every medium is numbered from 0 to 4. Spacer stands for h-BN, FL, Mica and Au. The color of the arrows indicates reflection/transmission in a different media and the transparency indicates the loss of intensity due to absorption after multiple reflections. (b) Schematic ray diagram to calculate the effective reflection at the spacer/SiO$_2$ interface. Every $r_i$ indicates a reflected ray from the 23 interface. The sum over all $r_i$ gives $r'$ (c) Schematic ray diagram to calculate the effective reflection at the MoS$_2$/spacer + SiO$_2$ interface. Note that the gray horizontal line between media 3 (SiO$_2$) and 4 (Si) indicates that media 3 and 4 are treated as one effective semi-infinite medium. Every reflection at the 23 interface is now treated as $r'$ and every reflection at the 12 interface is designated as $r_i''$. The sum over all $r_i''$ gives $r''$. (d) Schematic ray diagram of the absorption of the incoming laser beam by the MoS$_2$ layer. The black dots represent the points of absorption of the laser light at a depth $x$ in the MoS$_2$. Note that the gray lines at the 23 and 34 interfaces indicate that the stack of media 2 (spacer), 3 (SiO$_2$) and 4 (Si) are now treated as one effective semi-infinite medium. (e) Schematic ray diagram of the emission from a depth $x$ in the MoS$_2$ layer. The black dots represent the point of absorption of the laser light at a depth $x$ in the MoS$_2$. Note that the gray lines at the 23 and 34 interfaces indicate that the stack of media 2 (spacer), 3 (SiO$_2$) and 4 (Si) are now treated as one effective semi-infinite medium

The system is defined as schematically shown in Figure S9. Each media is numbered: 0 is air, 1 is $MoS_2$, 2 is either h-BN, FLG, Au or Mica, 3 is $SiO_2$ and 4 is Si. Each layer possess a complex index of refraction of the form $\mathbf{n}_i = n_i - ik_i$ where $n_i$ and $k_i$ are the real and imaginary part of the complex refraction index of the $i$-th layer. The Fresnel reflection ($r_{ij}$) and transmission ($t_{ij}$) coefficients are defined as follows:

$$r_{ij} = \frac{\mathbf{n_i} - \mathbf{n_j}}{\mathbf{n_i} + \mathbf{n_j}} \text{ and } t_{ij} = \frac{2\mathbf{n_i}}{\mathbf{n_i} + \mathbf{n_j}}$$

for a ray travelling from medium $i$ to medium $j$ and impinging on the $ij$ interface [8].

The following relations hold: $r_{ij} = -r_{ji}$ and $t_{ij}t_{ji} - r_{ij}r_{ji} = 1$

Travelling through any medium, the beam will acquire a geometric phase difference defined as $\beta_i = 2\pi \mathbf{n_i} \frac{d_i}{\lambda}$ with $n \geq 1$. Moreover, at a point $x$ in the depth of layer 1, the phase difference can be defined as $\beta_x = 2\pi \mathbf{n_1} \frac{x}{\lambda}$. Given the complexity of the system (Figure S9a), we will define an effective semi-infinite medium composed of media 3 and 4 (Figure S9b) and then define another effective semi-infinite medium composed of media 2 and the first effective medium (Figure S6c). The relevant quantity to calculate is the effective reflection coefficient of each effective medium ($r'$ and $r''$) taking into account multiple reflections. To calculate the first effective medium, we focus our attention at the 23 interface. Each component of the total reflection ($r_i$) is defined as :

$$r_1 = r_{23}$$

$$r_2 = t_{23}e^{-i\beta_3} \cdot r_{34}e^{-i\beta_3} \cdot t_{32} = t_{23}t_{32}t_{34}e^{-2i\beta_3}$$

$$r_3 = t_{23}e^{-i\beta_3} \cdot r_{34}e^{-i\beta_3} \cdot r_{32}e^{-i\beta_3} \cdot r_{34}e^{-i\beta_3} \cdot t_{32} = r_2(r_{32}r_{34}e^{-2i\beta_3})^1$$

…

$$r_n = r_2(r_{32}r_{34}e^{-2i\beta_3})^{n-2}$$

The total reflection is the sum over all $r_n$ and, taking into account the relationships between the transmission and reflection coefficients expressed before, it can be reduced to a geometric sum with result

$$r' = \frac{r_{23} + r_{34}e^{-2i\beta_3}}{1 + r_{23}r_{34}e^{-2i\beta_3}}. \tag{1}$$

The same procedure can now be applied to the interface between media 1 and 2 (see Figure S9c). The derivation above holds with the following substitutions: $r_{23} \to r_{12}$, $r_{34} \to r'$ and $\beta_3 \to \beta_2$ and we obtain:

$$r'' = \frac{r_{12} + r'e^{-2i\beta_2}}{1 + r_{12}r'e^{-2i\beta_2}} \tag{2}$$

which represents the total reflection at the 12 interface taking into account multiple reflection at the 34 and 23 interfaces. We note that it is possible to obtain the same result by employing the transfer matrix formalism.

With this coefficient we can proceed to the calculation of the absorption of light at a depth $x$ in the MoS$_2$ layer (Figure S6d). We have the following components

$$abs_1 = t_{01}e^{-i\beta_x}$$

$$abs_2 = t_{01}e^{-i\beta_1} \cdot r''e^{-1(\beta_1-\beta_x)} = t_{01}r''e^{(i2\beta_1-\beta_x)}$$

$$abs_3 = t_{01}e^{-i\beta_1} \cdot r'' e^{-i\beta_1} \cdot r_{10}e^{-i\beta_x} = abs_1(r_{10}r''e^{-i2\beta_1})$$

$$abs_4 = t_{01}e^{-i\beta_1} \cdot r'' e^{-i\beta_1} \cdot r_{10}e^{-i\beta_1} \cdot r'' e^{-1(\beta_1-\beta_x)} = abs_2(r_{10}r''e^{-2i\beta_2})$$

…

$$abs_{2n+1} = abs_1(r_{10}r''e^{-2i\beta_2})^n$$

$$abs_{2n+2} = abs_2(r_{10}r''e^{-2i\beta_2})^n$$

The total amount of absorbed light is given by the sum over all the $n$ terms and results in[8]

$$\boxed{F_{abs} = t_{01} \cdot \frac{e^{-i\beta_x} + r''e^{-i(2\beta_1-\beta_x)}}{1+r''r_{01}e^{-i2\beta_1}}} \tag{3}$$

The same derivation holds for the total amplitude of emitted light from depth $x$ in the MoS$_2$ (figure S6e) with the appropriate substitution $t_{01} \to t_{10}$ [8]:

$$\boxed{F_{sc} = t_{10} \cdot \frac{e^{-i\beta_x} + r''e^{-i(2\beta_1-\beta_x)}}{1+r''r_{01}e^{-i2\beta_1}}} \tag{4}$$

Note that the complex index of refractions are evaluated at the excitation wavelength for $F_{abs}$ and at the emission wavelengths for $F_{sc}$

Therefore, the intensity of the emitted light from the MoS$_2$ at a wavelength $\lambda$ is given by [8]

$$\boxed{I_{MoS_2}(\lambda) = \int_0^{d_1} |F_{abs}(\lambda_0,x) \cdot F_{sc}(\lambda,x)|^2 \, dx} \tag{5}$$

Expression 5 is valid for a general number of staked media. Hence, it can be used to calculate the intensity of emission in the case of 1L MoS$_2$ on top of SiO$_2$/Si (4 media), on top of Gel-Film (3 media) and suspended (3 media, Air / MoS$_2$ / Air).

**Calculation of the enhancement factor.**

In the above section, we presented the derivation and results of the model used to calculate possible interference effects due to the substrate geometry. In this section, we will use those results to calculate the enhancement of the emission of MoS$_2$ due to substrate-specific optical interference effects. These effects are well known for graphene [9-11] and can be evaluated by calculating an enhancement factor ($\Gamma^{-1}$) as the ratio of the emission intensity for 1L in the freestanding case and on the substrate and the emission:

$$\Gamma^{-1} = \frac{I_{MoS_2}^{\text{freestanding}}}{I_{MoS_2}^{\text{on substrate}}}$$

where $I_{MoS_2}^{\text{on substrate}}$ is the calculated emission intensity of 1L MoS$_2$ on a substrate and $I_{MoS_2}^{\text{freestanding}}$ is the emission intensity of 1L MoS$_2$ in a freestanding case.

Therefore, $\Gamma^{-1}$ describes the possible enhancement of the emission intensity due to interference effects within the substrate with respect to the freestanding condition. Knowing this enhancement factor allow us to compare the measured emission spectra from different substrates. To do so, we use $\Gamma^{-1}$ as a scaling factor to scale the data acquired on the substrate to the freestanding condition in the following way:

$$I_{MoS_2}^{\text{freestanding}} = MI_{MoS_2}^{\text{on substrate}} \cdot \Gamma^{-1}$$

where $MI_{MoS_2}^{\text{on substrate}}$ is the measured intensity on a substrate, $\Gamma^{-1}$ is the enhancement factor and $I_{MoS_2}^{\text{freestanding}}$ is the rescaled data to freestanding condition. By multiplying the measured data ($MI_{MoS_2}^{\text{on substrate}}$) with the enhancement factor ($\Gamma^{-1}$) we take into account the effect of the interference within the substrate and rescale the data to the freestanding condition ($I_{MoS_2}^{\text{freestanding}}$).

**Effect of spacer height on $\Gamma$, Raman and PL spectra of 1L MoS$_2$.**

In the previous sections we developed a model to take into account the interference effects within a substrate and calculate the enhancement factor with respect to a freestanding MoS$_2$ layer. Both of those results are dependent on the geometry of the sample and especially the thickness of the different layers. In this section we will elucidate the effect of the thickness on the model and measured data.

Since most of the measurement are done on substrates comprising a stack of 285 nm SiO$_2$ on Si wafer, we first study the effect of a small variation of the height of the SiO$_2$ layer on the enhancement factor and Raman and PL spectra.

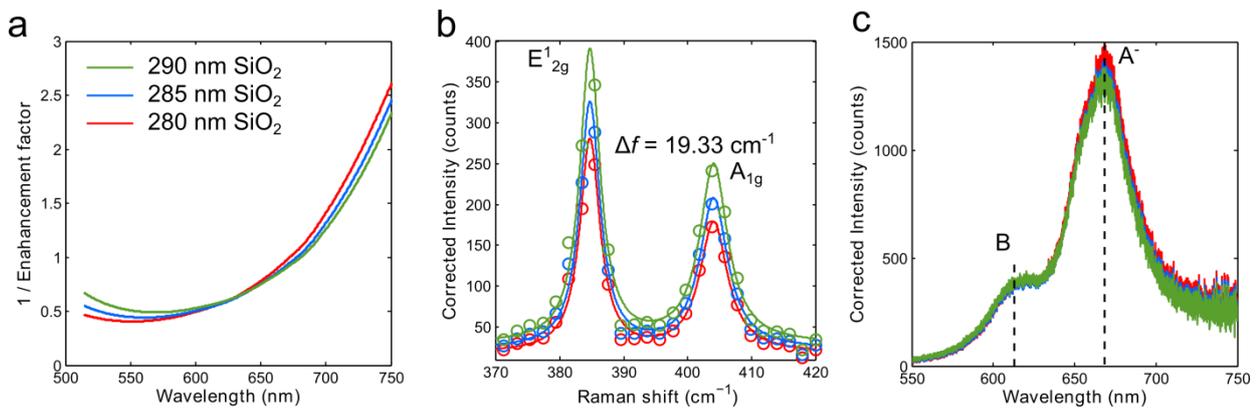

**Figure S10.** (a) Inverse of the Enhancement Factor ($\Gamma^{-1}$) vs wavelength for 290 nm SiO$_2$ (green line), 285 nm SiO$_2$ (blue line), 280 nm SiO$_2$ (red line). (b) Corrected intensity of the Raman part of the spectrum against Raman shift (cm$^{-1}$) for 290 nm SiO$_2$ (green line and dots), 285 nm SiO$_2$ (blue line and dots), 280 nm SiO$_2$ (red line and dots). The E$^1_{2g}$ and A$_{1g}$ mode are identified and their difference ($\Delta f$) is 19.33 cm$^{-1}$. (c) Corrected intensity of the PL part of the spectrum against wavelength for 290 nm SiO$_2$ (green line and dots), 285 nm

SiO₂ (blue line and dots), 280 nm SiO₂ (red line and dots).

Figure S10 shows the inverse of the enhancement factor (a) and effect of the effect on the Raman (b) and PL (c) of 1L MoS₂ of 280 nm, 285 nm and 290 nm of SiO₂/Si wafer substrate. It is evident that a small variation on the silicon oxide height does not significantly influence the data. The intensity of the Raman peaks can be enhanced by a maximum a factor 1.4 and the intensity in the emission from the A⁻ exciton is barely affected. It is interesting to note that the Raman part of the spectrum has the opposite trend with respect to the PL part: increasing the oxide thickness increases the Raman signal, while decreases the PL signal from the A⁻ exciton. Note that the wavelength of emission (both for Raman and PL) is not affected by the oxide thickness. Therefore, we can conclude that a small variation in the SiO₂ thickness has a negligible effect on the emission spectra of 1L MoS₂.

We can now study the effect of small height variation on the spacer height, keeping the SiO₂ thickness fixed to 285nm. The studied spacer height values are close to the spacer height in the measured structures so that we can address the effect on the spectra of a small uncertainty in the determination of the spacer height.

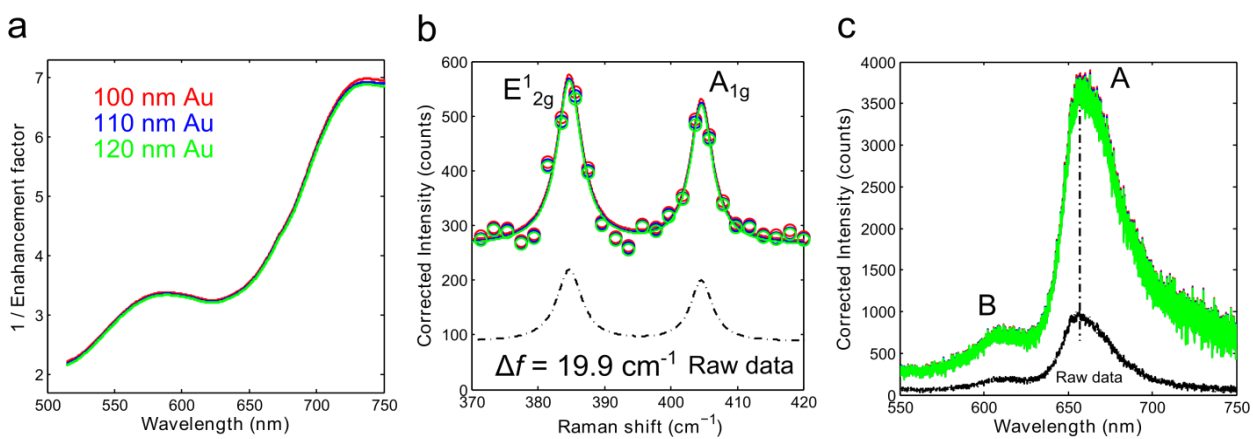

**Figure S11.** (a) Inverse of the Enhancement Factor ($\Gamma^{-1}$) vs wavelength for 120 nm Au (green line), 110 nm Au (blue line), 100 nm h-BN (red line). (b) Corrected intensity of the Raman part of the spectrum against Raman shift (cm$^{-1}$) for 120 nm Au (green line), 110 nm Au (blue line), 100 nm h-BN (red line). The dashed-dotted black line represents the raw data. The E$^1_{2g}$ and A$_{1g}$ mode are identified and their difference ($\Delta f$) is 19.9 cm$^{-1}$. (c) Corrected intensity of the PL part of the spectrum against wavelength for 300 nm h-BN (green line), 290 nm h-BN (blue line), 280 nm h-BN (red line). The dashed black line represents the raw data. The dashed-dotted line indicates the position of the A peak.

Figure S11 shows the inverse of the enhancement factor (a) and effect of the effect on the Raman (b) and PL (c) of 1L MoS$_2$ on 120 nm, 110 nm and 100 nm of Au. From Figure S10a it is evident that $\Gamma^{-1}$ has a strong dependence on the wavelength but negligible on the thickness of the Au spacer. The negligible dependence on the height of the Au spacer is reflected in the Raman (Figure S11b) and PL (Figure S11c) part of the emission that do not show any variations with respect the variations in the height of the h-BN spacer. Converted into a freestanding condition, the Au substrate provides a factor ~7 enhancement of the raw data in the PL part of the spectrum.

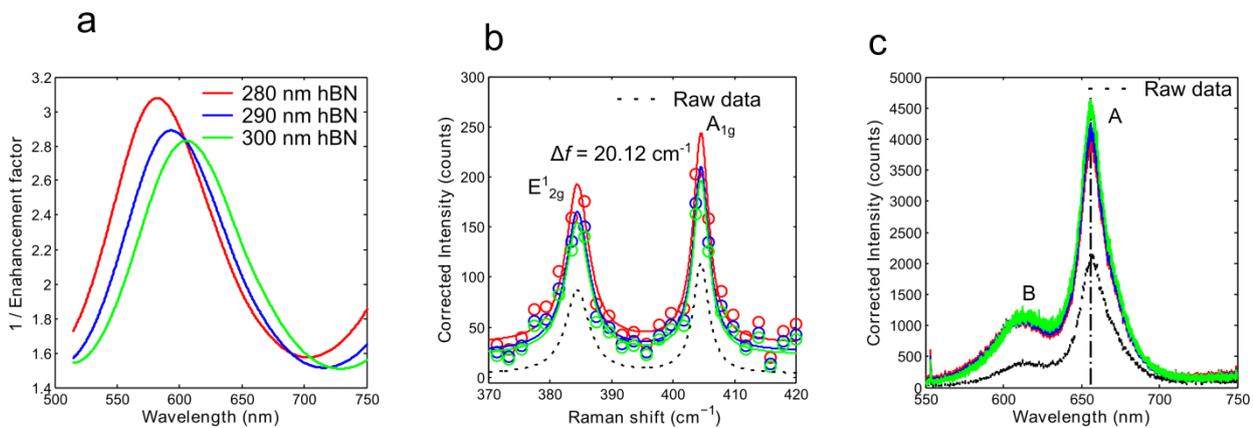

**Figure S12.** (a) Inverse of the Enhancement Factor ($\Gamma^{-1}$) vs wavelength for 300 nm h-BN (green line), 290 nm h-BN (blue line), 280 nm h-BN (red line). (b) Corrected intensity of the Raman part of the spectrum against Raman shift (cm$^{-1}$) for 2 for 300 nm h-BN (green line), 290 nm h-BN (blue line), 280 nm h-BN (red line). The dashed black line represents the raw data. The E$^1_{2g}$ and A$_{1g}$ mode are identified and their difference ($\Delta f$) is 20.12 cm$^{-1}$. (c) Corrected intensity of the PL part of the spectrum against wavelength for 300 nm h-BN (green line), 290 nm h-BN (blue line), 280 nm h-BN (red line). The dashed black line represents the raw data. The dashed-dotted line indicates the position of the A peak.

Figure S12 shows the inverse of the enhancement factor (a) and effect on the Raman (b) and PL (c) of 1L MoS$_2$

on 300 nm, 290 nm and 280 nm of h-BN. From Figure S10a it is evident that $\Gamma^{-1}$ depends on the wavelength and it shows variations of a factor ~2 across the whole spectral range. This is reflected in the Raman (Figure S12b) and PL (Figure S12c) part of the emission that also do not show significant variations with respect to the variations in the height of the h-BN spacer, especially in the position of the emission peaks. The optical interference provides a factor ~2 enhancement of the raw data in the PL part of the spectrum, when converted back to a freestanding condition. Note the opposite trend for Raman and PL emission over the height variation of the h-BN spacer.

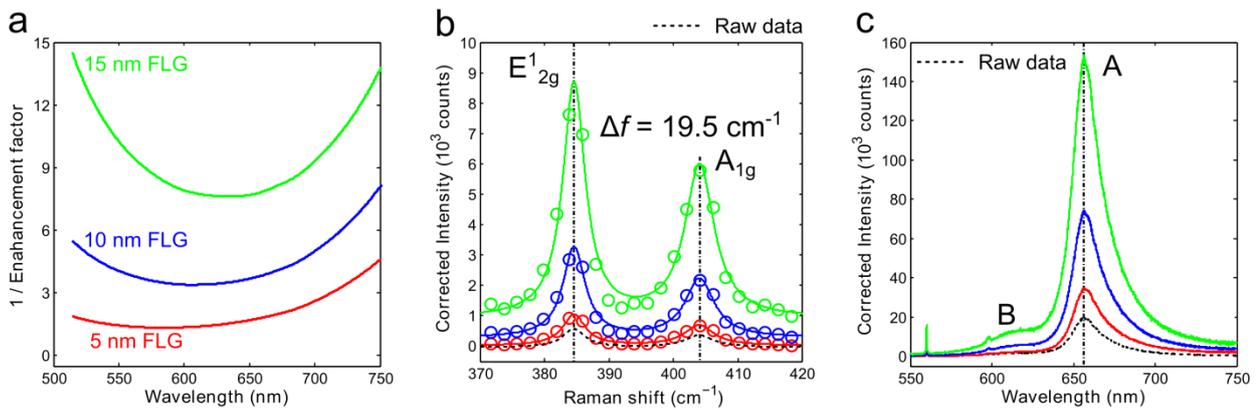

**Figure S13.** (a) Inverse of the Enhancement Factor ($\Gamma^{-1}$) vs wavelength for 15 nm FLG (green line), 10 nm FLG (blue line), 5 nm FLG (red line). (b) Corrected intensity of the Raman part of the spectrum against Raman shift (cm$^{-1}$) for 2 for 15 nm FLG (green line), 10 nm FLG (blue line), 5 nm FLG (red line). The dashed black line represents the raw data. The $E^1_{2g}$ and $A_{1g}$ mode are identified and their difference ($\Delta f$) is 19.5 cm$^{-1}$. (c) Corrected intensity of the PL part of the spectrum against wavelength 15 nm FLG (green line), 10 nm FLG (blue line), 5 nm FLG (red line). The dashed black line represents the raw data. The dashed-dotted line indicates the position of the A peak.

Figure S13 shows the inverse of the enhancement factor (a) and effect on the Raman (b) and PL (c) emission of 1L MoS$_2$ on 15 nm, 10 nm and 5 nm of FLG. From Figure S10a it is evident that $\Gamma^{-1}$ is strongly dependent on the thickness of the FLG flake, showing variations of a factor ~6 from 5 nm FLG to 15 nm FLG. This is reflected in the Raman (Figure S13b) and PL (Figure S13c) part of the emission that also clearly show significant variations with respect to the variations in the height of the FLG spacer: the counts range from ~38 $10^3$ to ~160 $10^3$ going from 5nm to 15nm FLG. The position of the emission peak is not affected by the height of the spacer. Note the common trend for Raman and PL emission intensity over the height variation of the FLG spacer.

**Fit to 1L MoS$_2$ PL data**

In this section we show the fit to the experimental PL data on 1L MoS$_2$. The fits are composed by 3 Lorentzians and 1 linear background.

Figure S14 shows the both the full fit (red solid line) superimposed to the data (blue solid line) and the functions that compose the fit (solid lines: blue for the B exciton, green for the A exciton, red for the A-exciton and black for the linear background).

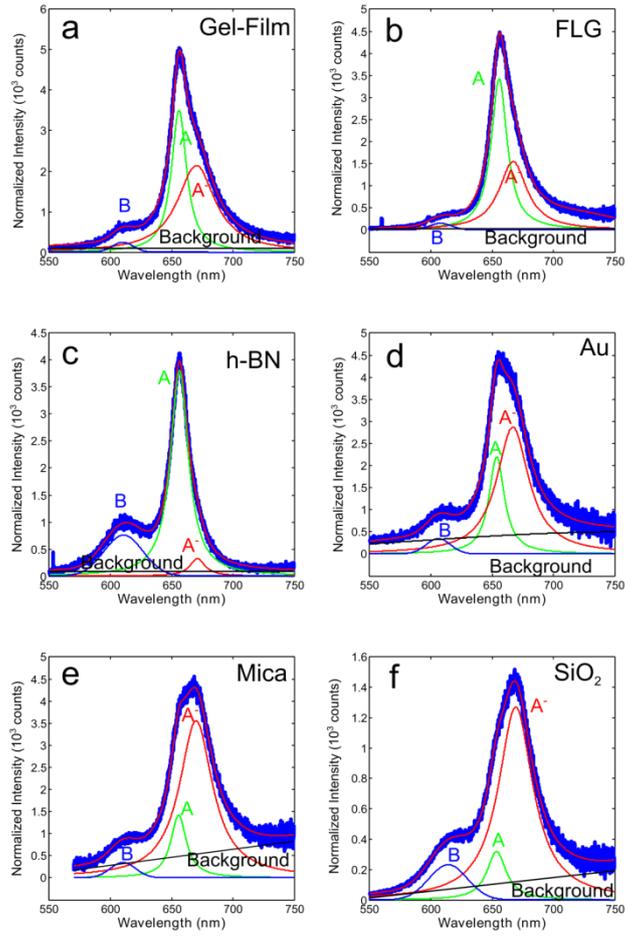

**Figure S14.** Fits to the PL data to 1L MoS$_2$ on gel-film (a), FLG (b), h-BN (c), Au (d), Mica (e), SiO$_2$ (f). the fitting functions are identified in the figure.

**Mass action model for trions**

In this section we will develop the mass action model for the excitons and trion population in 1L MoS$_2$, which can be considered as a 2D electron gas. This model is based on the dynamic equilibrium between neutral excitons (*A*), free electrons (*e$_{free}$*) and negatively charged excitons or trions (*A$^-$*) [12]. This dynamic equilibrium can be expressed as

$$A + e_{free} = A^-$$

Thus, a neutral exciton can be seen as a binding site for a free electrons to generate charged excitons. It is possible to define the ratio between the density of A ($N_A$) and A⁻ ($N_{A^-}$) as

$$\frac{N_A}{N_{A^-}} = \frac{\beta_A}{\beta_{A^-}} \cdot \exp(-\frac{\mu+\epsilon}{k_B T})$$

Where $\beta_A$ is the number of degenerate spin configurations of neutral excitons and $\beta_{A^-}$ is by the number of degenerate spin configurations of charged excitons, $\mu$ is the chemical potential of electrons, $\varepsilon$ is the trion binding energy (18 meV [13]), $k_B$ is the Boltzmann constant and $T$ is the temperature. Neutral excitons have 2x4 degenerate configurations, so $\beta_A = 8$. Trions have $\beta_{A^-} = 2$ (due to spin coupling of the two electrons) [12]. We can define the chemical potential as $\mu = k_B T \ln\left(\exp\left(\frac{E_F}{k_B T}\right) - 1\right)$, where $E_F = \frac{\pi \hbar^2 N_e}{m^*}$ is the Fermi energy of the system, $\hbar$ is the reduced Planck's constant, $N_e$ is the free electron density and $m^*$ is the effective mass. Note that $N_e$ is composed of the natural doping and the photo-generated carriers under illumination in the MoS₂. The effective mass is defined as $m^* = \frac{m_A \cdot m_e}{m_{A^-}} = \frac{0.8 \cdot m_0 \cdot 0.35 \cdot m_0}{1.15 \cdot m_0}$, where $m_A$ is the reduced mass of the neutral exciton, $m_{A^-}$ is the reduced mass of the trion and $m_e$ of the electron and $m_0$ is the free electron mass [13].

Coupled to the mass action model, we consider a three level model and corresponding rate equations to define the emission intensities of the excitons. The three levels are: the ground level, the neutral exciton and the charged exciton. The radiative decay rate from the neutral exciton to ground and from the trion to ground are $\gamma_A$ and $\gamma_{A^-}$ respectively. Therefore the emission intensities can be expressed as $I_A = \tau_A \cdot N_A$ for the neutral exciton and $I_{A^-} = \tau_{A^-} \cdot N_{A^-}$ for the charged excitons. From the above definitions, it is possible to calculate the ratio between the emission of the trions with respect to the total emission (from the A excitonic species)

$$\gamma = \frac{I_{A^-}}{I_A + I_{A^-}} = \frac{\tau_{A^-} \cdot N_{A^-}}{\tau_A \cdot N_A + \tau_{A^-} \cdot N_{A^-}} = \frac{1}{\frac{\tau_A}{\tau_{A^-}} \frac{N_A}{N_{A^-}} + 1}$$

The only unknown is the ratio between the decay rates. This has been recently studied by Mouri *et al*.[14] and the value is around ~6.6.

**Calculation for model in Figure 4d**

In this section, we describe the formalism employed to calculate the theoretical relationship displayed in Figure 4d in the main text. Given the above model, we need to identify a relationship between the doping level and the frequency of the $A_{1g}$ Raman mode of 1L MoS$_2$. From the study of Chakraborty *et al* [15], we can identify this relationship in the form: $\frac{\Delta n}{\Delta \omega} \sim -0.22 \times 10^{13}$ cm$^{-2}$ / cm$^{-1}$. The gray dashed line in Figure 4d is obtained with a starting free electron density for the 1L MoS$_2$ on SiO$_2$ $\sim 7 \times 10^{12}$, reducing at the above rate. Note that this density is the sum of the intrinsic doping level in the MoS$_2$ and the photoexcited electrons generated in the MoS$_2$ under illumination. This value compares well with recent measurements of the intrinsic doping level in 1L MoS$_2$ on SiO$_2$ made with electrostatic AFM techniques that delivered a value of $\sim 5 \times 10^{12}$ cm$^{-2}$ [16].

**Spatially resolved PL maps**

In Figure S15-S18, we show the representative selection of the PL maps measured on the studied MoS$_2$-based structures. In panels(a) of Figures S1-S5, we present optical micrographs of the studied structures. In panels(b) of Figures S1-S5, we show the spatial map of the peak emission wavelength. In panels(c) of Figures S1-S5, we show spatially resolved integrated intensity.

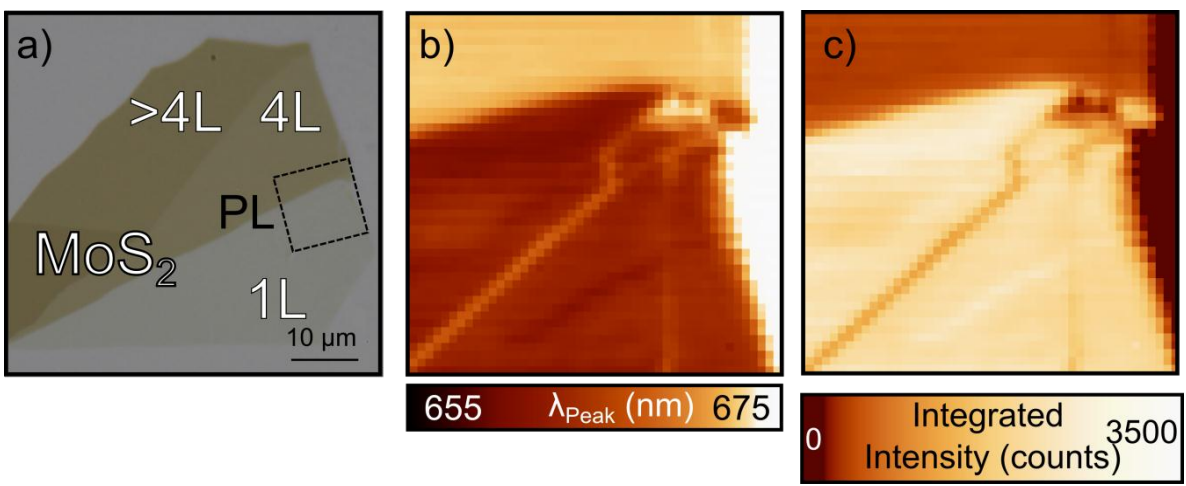

**Figure S15.** (a) optical micrograph of few layer MoS$_2$ on gel-film. (b) Peak PL emission wavelength of the MoS$_2$/Gel-Film® device in Figure S3. (c) Integrated PL emission of the MoS$_2$/Gel-Film® device in Figure S3.

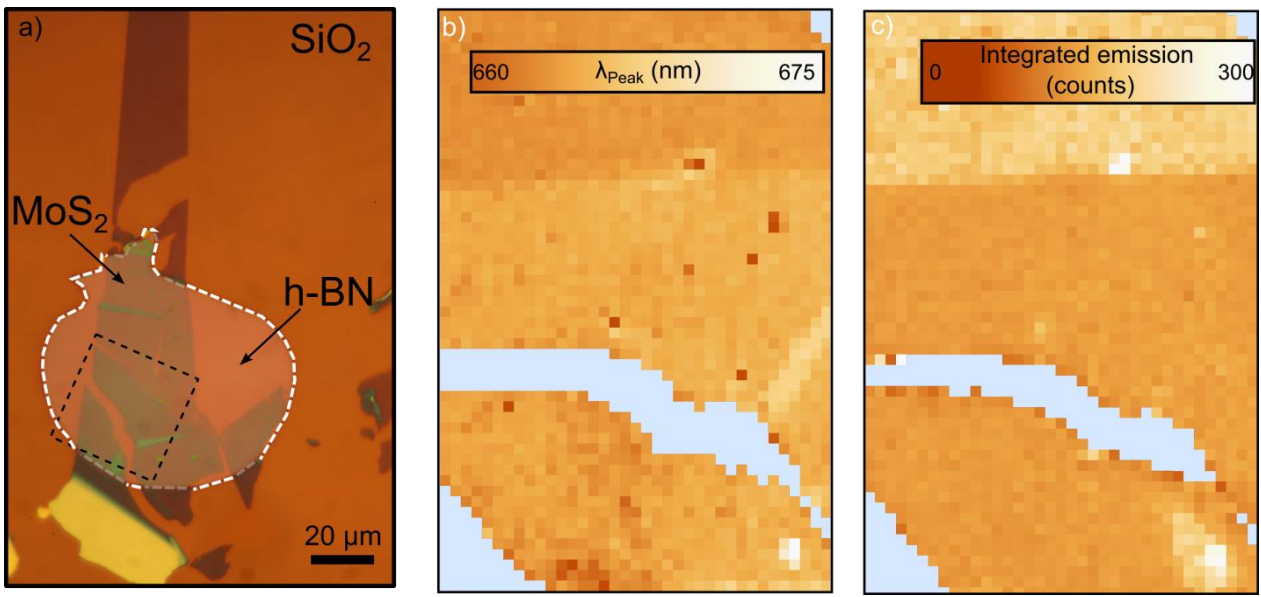

**Figure S16**. (a) optical micrograph of few layer MoS$_2$ on h-BN (b) Peak PL emission wavelength of the MoS2/h-BN device in Figure S4. (c) Integrated PL emission of the MoS2/Gel-Film® device in Figure S4

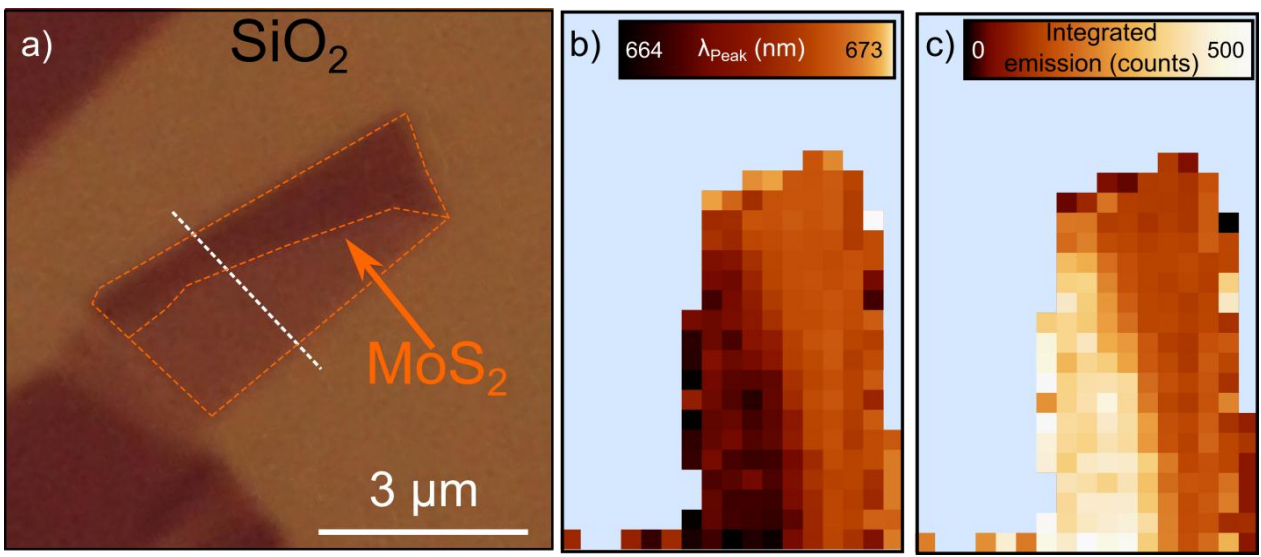

**Figure S17.** (a) optical micrograph of few layer MoS$_2$ on SiO$_2$ (b) Peak PL emission wavelength of the MoS2/SiO2 device in Figure S5. (c) Integrated PL emission of the MoS$_2$/SiO$_2$ device in Figure S5

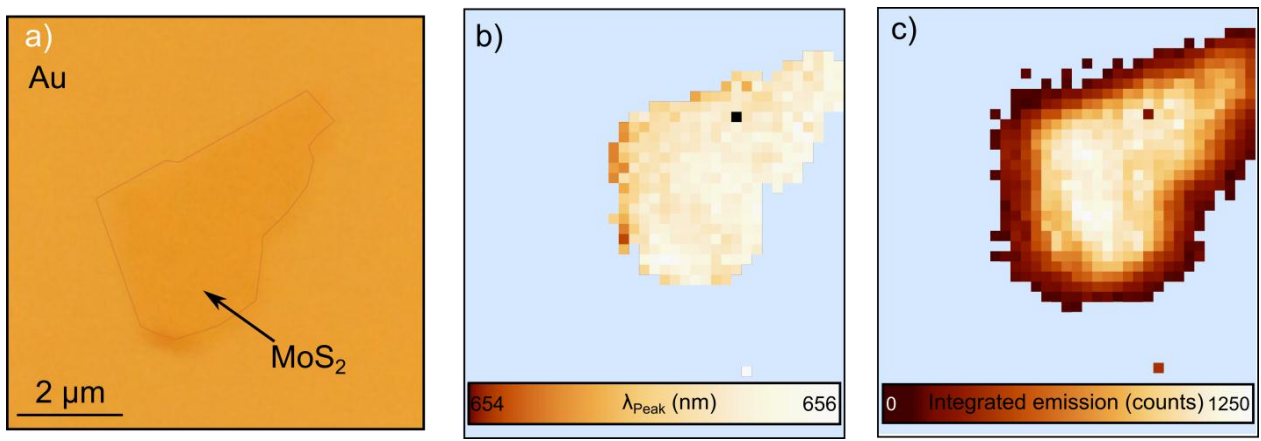

**Figure S18.** (a) optical micrograph of few layer MoS$_2$ on Au (b) Peak PL emission wavelength of the MoS$_2$/Au device in Figure S1. (c) Integrated PL emission of the MoS$_2$/Au device in Figure S1

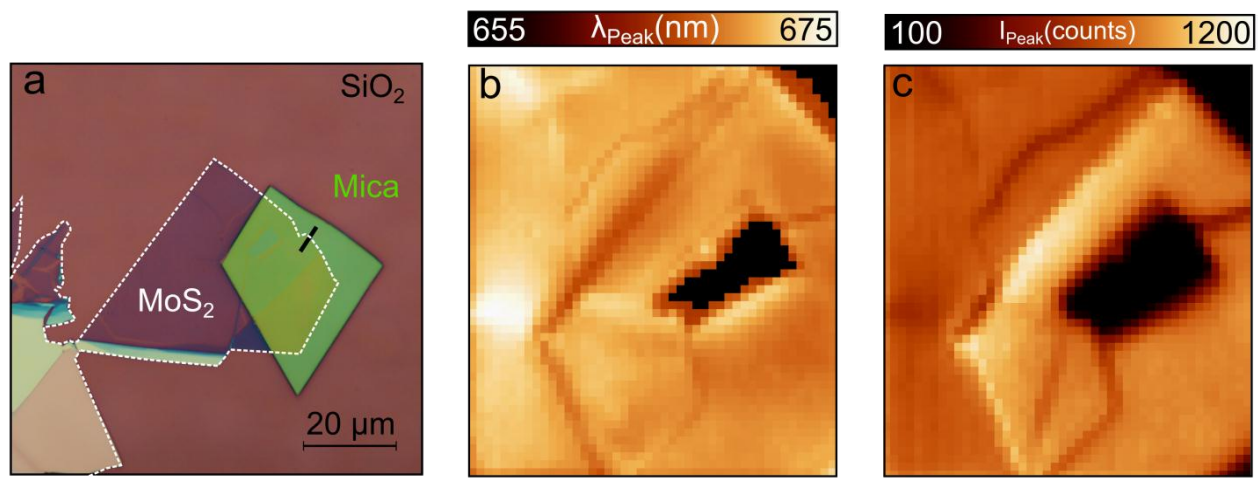

**Figure S19.** a) optical micrograph of few layer MoS$_2$ on Mica (b) Peak PL emission wavelength of the MoS$_2$/Au device in Figure S1. (c) Integrated PL emission of the MoS$_2$/Au device in Figure S1

**Determination of the FLG edge.**

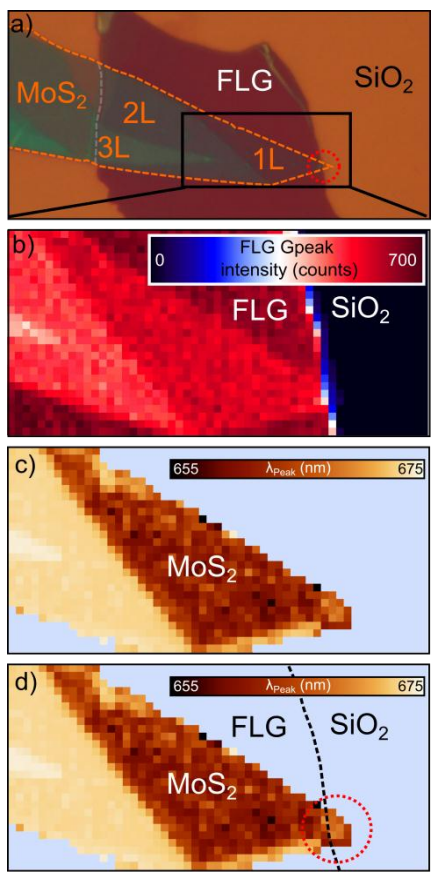

**Figure S20.** (a) Optical micrograph of the MoS$_2$/FLG/SiO$_2$ heterostrucutre as in Figure 1 and 2 in the main text. (b) Intensity of the G-peak of the FLG flake. (c) Peak emission wavelength from the MoS$_2$ flake. (d) Peak emission wavelength from the MoS$_2$ flake with the FLG edges identified by superposition of panel (c).

Figure S20 shows the procedure for determining the edgs of the FLG flake of the MoS$_2$/FLg heterostrucutre shown in Figure 2 in the main text. Figure S20a shows an optical micrograph of the device where it is clearly possible to identify the edges of the FLG and MoS$_2$ flakes. Layer determination is done via Raman spectroscopy (see main text Figure 1). Figure S20b plots a colormap of the intensity of the G peak. From this colormap, it is possible to distinguish the FLG from the SiO$_2$ background. Note the variation of the intensity of the G-peak. Figure S20c plots a colormap of the wavelength of the peak PL emission from the MoS$_2$ flake. From this colormap, it is possible to isolate the MoS$_2$ flake from the backsground, since it is the only emitter in the 650 – 680 nm wavelength range. Figure S18d is obtained by superposition of Figure S20b and c. The edge of the FLG flake is determined with FigureS20b and represented by the dashed white line. It is thus possible to distinguish which part of the MoS$_2$ flake are supported by the FLG and the SiO$_2$ and correlated their emission properties to the substrate. It is clear that the intensity of the G peak is related to number of MoS$_2$ layers on top of the FLG flake. This is an extra tool to confirm the number of MoS$_2$ layers composing the heterostrucure.

___________

Address correspondence to Michele Buscema m.buscema@tudelft.nl. Andres Castellanos-Gomez: a.castellanosgomez@tudelft.nl